\documentclass[aps,prx,reprint,superscriptaddress]{revtex4-2}

\usepackage{graphicx}
\usepackage{xcolor} % Used to create different colors for every author
\usepackage{multirow}
\usepackage{amsmath}
\usepackage{amsfonts}
\usepackage{hyperref}
\usepackage{float}
\usepackage{lipsum}
\usepackage[version=4]{mhchem}
\usepackage{placeins}
\usepackage{seqsplit}

\usepackage{booktabs}

\usepackage{bibunits}
\defaultbibliographystyle{achemso}
\defaultbibliography{refs}
\begin{document}

\author{Moritz Sallermann}
\affiliation
{Science Institute and Faculty of Physical Sciences, University of Iceland, VR-III, 107 Reykjav\'{\i}k, Iceland}
\affiliation
{Departamento de Qu\'{\i}mica F\'{\i}sica,
Facultad de Ciencias Qu\'{\i}micas, Universidad Complutense de Madrid,
28040 Madrid, Spain.}

\author{Amrita Goswami}
\affiliation
{Science Institute and Faculty of Physical Sciences, University of Iceland, VR-III, 107 Reykjav\'{\i}k, Iceland}
\affiliation
{Departamento de Qu\'{\i}mica F\'{\i}sica,
Facultad de Ciencias Qu\'{\i}micas, Universidad Complutense de Madrid,
28040 Madrid, Spain.}

\author{Rosana Collepardo-Guevara}
\affiliation
{Yusuf Hamied Department of Chemistry, University of Cambridge Lensfield Road, Cambridge CB2 1EW, United Kingdom}
\affiliation
{Department of Genetics, University of Cambridge, Cambridge CB2 3EH, UK.}

\author{Alberto Ocana}
\affiliation
{Multidisciplinary Institute, University Complutense of Madrid, Paseo Juan XXIII, 1, Madrid 28040, Spain}
\affiliation
{Phasica Biosciences S.L, Calle Velázquez, 27, 28001 Madrid, Spain}
\affiliation{Experimental Therapeutics CRIS Cancer Unit, Hospital Clinico San Carlos and IdSCC and CIBERONC, Madrid, Spain.}
\affiliation{
Cátedra INTHEOS-START-CEU. START Madrid-Fundación Jiménez Díaz (FJD) Early Phase Program, Fundación Jiménez Díaz Hospital, Madrid, Spain.}

\author{Hannes J\'{o}nsson}
\affiliation
{Science Institute and Faculty of Physical Sciences, University of Iceland, VR-III, 107 Reykjav\'{\i}k, Iceland}

\author{Elvar \"O. J\'{o}nsson}
\email[]{elvarorn@hi.is}
\affiliation
{Science Institute and Faculty of Physical Sciences, University of Iceland, VR-III, 107 Reykjav\'{\i}k, Iceland}

\author{Jorge R. Espinosa}
\email[]{jorgerene@ucm.es}
\affiliation
{Departamento de Qu\'{\i}mica F\'{\i}sica,
Facultad de Ciencias Qu\'{\i}micas, Universidad Complutense de Madrid,
28040 Madrid, Spain.}
\affiliation
{Yusuf Hamied Department of Chemistry, University of Cambridge Lensfield Road, Cambridge CB2 1EW, United Kingdom}
\affiliation
{Phasica Biosciences S.L, Calle Velázquez, 27, 28001 Madrid, Spain}
\affiliation
{Multidisciplinary Institute, University Complutense of Madrid, Paseo Juan XXIII, 1, Madrid 28040, Spain}

%%%%%%%%%%%%%%%%%%%%%%%%%%%%%%%%%%%%%%%%%%%%%%%%%%%%%%%%%%%%%%%%%%%%%

\title
{
ChemFit: A framework for automated high-dimensional model parameter optimization
}

%%%%%%%%%%%%%%%%%%%%%%%%%%%%%%%%%%%%%%%%%%%%%%%%%%%%%%%%%%%%%%%%%%%%%
% should be upto 200 words
% currently 235 words
\begin{abstract}
The parameterization of simulation-based models is a central yet laborious task in computational chemistry and physics, often driven by human intuition and manual iteration. Automating this task necessitates the definition of suitable objective functions, which tend to be expensive to evaluate, noisy, non-differentiable, or composed of heterogeneous contributions originating from separate sets of simulations. Gradient-free and black-box optimization algorithms are powerful tools which are particularly well-suited to minimizing such objective functions. Here, we introduce \texttt{ChemFit}, a flexible Python framework for the definition, composition, and massively concurrent evaluation of simulation-based objective functions, which is designed to operate in conjunction with these algorithms. We demonstrate the broad applicability of this approach by using \texttt{ChemFit} for three representative examples of increasing complexity and real-world relevance. First, we obtain the parameters of the Lennard-Jones potential for liquid argon from experimental measurements of the density. Second, we parameterize a polarizable and flexible potential energy function to reproduce the structure of small \ce{H2O} clusters obtained from density functional theory calculations. Finally, we tune a small subset of the parameters of a residue-level coarse-grained protein force-field, with the goal to reproduce the experimental critical solution temperature of the low complexity domain of the wild-type hnRNPA1 sequence and an arginine-enriched mutant of this protein. hnRNPA1 is an RNA-binding protein linked to amyotrophic lateral sclerosis. Together, these examples illustrate how \texttt{ChemFit} enables scalable, reproducible, and optimizer-agnostic parameter fitting for broadly applicable multiscale models.
\end{abstract} 

\maketitle

%%%%%%%%%%%%%%%%%%%%%%%%%%%%%%%%%%%%%%%%%%%%%%%%%%%%%%%%%%%%%%%%%%%%%
%% Introduction
%%%%%%%%%%%%%%%%%%%%%%%%%%%%%%%%%%%%%%%%%%%%%%%%%%%%%%%%%%%%%%%%%%%%%
\begin{bibunit}

%%%%%%%%%%%%%%%%%%%%%%%%%%%%%%%%%%%%%%%%%%%%%%%%%%%%%%%%%%%%%%%%%%%%%
%% Introduction
%%%%%%%%%%%%%%%%%%%%%%%%%%%%%%%%%%%%%%%%%%%%%%%%%%%%%%%%%%%%%%%%%%%%%
\section{Main}

In contemporary computational chemistry, physics, and related theoretical disciplines, the calibration of model parameters against experimental measurements or high-level theoretical calculations is a fundamental component of model development and validation. Accurate parameterization is essential for addressing major scientific and technological challenges across multiple models, length and time scales. Representative examples include the fitting of interatomic potentials and molecular force fields~\cite{Harrison2018,Mueser2022,Jorgensen1996,Hornak2006,leontyev10a,kirby2019charge,cruces2024building,jorge2019dielectric,le2020molecular,wikfeldt_2013_transferablea}, the parameterization of coarse-grained models~\cite{Noid2013,Brini2013,Molinero2008,Joseph2021,Souza2021,Cao2024,kapcha2014simple,dignon2018sequence}, and the optimization of phenomenological models employed in geophysics or materials science~\cite{Hess2020,Fedotov2016,negro2013lava,cordonnier_2016_benchmarking}.

To paraphrase a quote by Gottfried Wilhelm Leibnitz, co-inventor of calculus and designer of an early computing machine: It is unworthy to lose hours in the labour of calculation which would safely be relegated to machines. In light of this, we aim to reduce the human effort in model parametrization. The first step to automating this process is the definition of an appropriate objective function that quantifies how well the model performs. These objective functions can range from obvious to highly complex multi-parameter functions, depending on the precise modelling task. However, besides the key step of definition, an objective function can exhibit characteristics that challenge traditional optimization approaches: it may be noisy due to finite sampling, non-differentiable due to discrete events or phase transitions, and composed of many heterogeneous contributions associated with independent data points or distinct physical conditions~\cite{Kolda2003,Wild2008}. As a result, gradient-based optimization methods are often inapplicable, while grid-based parameter sweeps scale exponentially with dimensionality and rapidly become computationally prohibitive~\cite{Rios2012}.

% so that the figure is on the same page as the Results
\begin{figure*}[t!]
    \includegraphics[width=\linewidth]{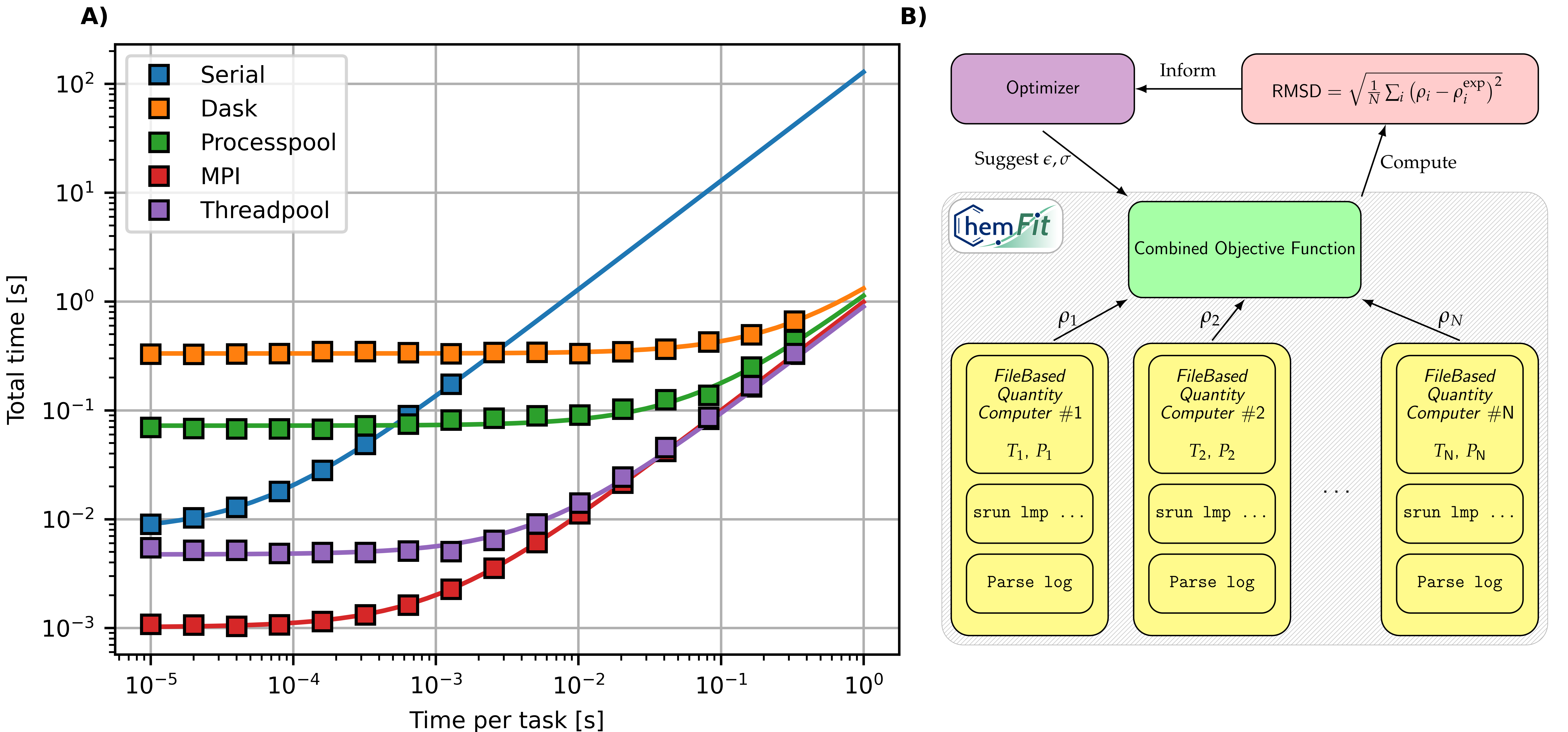}
    \caption{
    \textbf{A)} Total time taken to evaluate a combined objective function with ten individual terms ($y$-axis), versus the time it takes to evaluate each individual term ($x$-axis). Different colors correspond to different concurrency schemes. The solid lines serve as visual guides and are linear fits to the data points (evenly weighted in log-log space). The total number of available threads or processes, respectively, was 48.
    \textbf{B)} Flowchart depicting the use of \texttt{ChemFit} with LAMMPS to fit to the experimental argon densities. For each experimental data point, a \emph{FileBasedQuantityComputer} is created, which handles the creation of an appropriate LAMMPS input script, for each trial parameter set of $\sigma$ and $\varepsilon$. The \emph{FileBasedQuantityComputer} also uses an output parser to analyze the log files created by LAMMPS to obtain the averaged densities. 
    }
    \label{fig:chemfit_code}
\end{figure*}

Gradient-free and black-box optimization methods---including evolutionary strategies~\cite{Baeck1993,Rakshit2017}, stochastic search~\cite{lin2022gradient,Kozak2021}, and Bayesian optimization~\cite{Shahriari2016,Frazier2018,Wang2023,Binois2022,goswamiAdaptivePruningIncreased2025,snoek2012practical,tesei2021accurate}---offer a powerful alternative~\cite{Jones1998,Rios2012}. These approaches treat the objective function as an opaque mapping from parameters to scalar loss values and have found widespread use in fields ranging from hyperparameter optimization in machine-learning~\cite{Bischl2023,pmlr-v133-turner21a,snoek2012practical} to experimental design and control~\cite{Terayama2021, foster2019variational}. While, arguably, these algorithms are mature and ready-to-use, particularly for hyperparameter optimization in machine-learning~\cite{Bischl2023, Li2021}, their application in the context of computational physics and chemistry is complicated by the need to orchestrate large numbers of concurrent, heterogeneous simulations. In addition, it is necessary to parse output data, aggregate results across many independent terms, and exploit concurrency at multiple levels to efficiently utilize the available computational resources.

In this work, we present \texttt{ChemFit}, which focuses explicitly on the definition and evaluation of complex objective functions arising in simulation-based parameter fitting. \texttt{ChemFit} enables the practical application of gradient-free optimization methods to large and realistic scientific problems, without presuming any particular optimization algorithm or mode of execution. We interface the objective functions, defined via \texttt{ChemFit}, with gradient-free optimization libraries~\cite{optuna_2019, nevergrad, 2020SciPy-NMeth}, and illustrate the capabilities of this approach for model parametrization through three representative case studies. As a proof-of-concept, we first estimate the parameters of a Lennard-Jones model for liquid argon, by fitting the model to available experimental data. We deliberately start from parameters that are far away from the literature values and successfully recover parameters fully consistent with the literature. Secondly, we increase the complexity of the problem by parameterizing a potential energy function describing polarizable and flexible \ce{H2O} molecules to reproduce the geometry of small \ce{H2O} clusters, obtained from density functional theory (DFT) calculations as a reference. Multiple parameters in the model were optimized, and we obtain close agreement with the DFT energy for the same set of water clusters. Finally, we adjust a reduced set of parameters of a state-of-the-art residue-level coarse-grained (CG) protein model \cite{R.Tejedor2025} to improve the agreement with experimental estimates of the upper critical solution temperature for phase-separation. We showcase this example for two intrinsically disordered proteins (IDPs)---the low complexity domain of hnRNPA1 and an arginine enriched +7R mutated variant of this sequence. We show how \texttt{ChemFit} successfully lowers the critical solution temperature of the +7R variant towards the experimental value, while preserving the critical temperature of the wild type sequence, which already matched the experimental \textit{in vitro} values \cite{bremerDecipheringHowNaturally2022}. Owing to the fact that our case studies span a relatively wide scientific background, Sec.~\ref{sec:results_lj} to Sec.~\ref{sec:results_mpipi} provide small introductions for each example. Overall, these case studies illustrate how \texttt{ChemFit} enables scalable, multi-parameter, method-agnostic fitting applicable for a wide range of models, across different resolutions and scientific domains. 
%for broadly applicable models across different resolutions.  

%%%%%%%%%%%%%%%%%%%%%%%%%%%%%%%%%%%%%%%%%%%%%%%%%%%%%%%%%%%%%%%%%%%%%
%% Results
%%%%%%%%%%%%%%%%%%%%%%%%%%%%%%%%%%%%%%%%%%%%%%%%%%%%%%%%%%%%%%%%%%%%%
\section{Results}

\subsection{The ChemFit computational platform}

\texttt{ChemFit} is a Python package enabling the massively parallel evaluation and coordination of simulation codes for the purpose of defining objective functions. Such objective functions can, for example, be used with gradient-free optimization routines. Further, objective functions in \texttt{ChemFit} can make use of a concurrency-safe and powerful context and meta-data system to provide provenance information.
\texttt{ChemFit} provides utilities to interface with optimization algorithms, such as the ones provided by Nevergrad~ \cite{nevergrad}, Scipy~\cite{2020SciPy-NMeth} or Optuna~\cite{optuna_2019}.

Concurrency in \texttt{ChemFit} can be incorporated in interchangeable, problem-adjusted manners (see Sec.\ref{subsec:concurrency} for further details). It is compatible with an abstract ``Executor" interface, which encapsulates the access to computational resources and is in turn implemented by several other libraries. Besides being compatible with this flexible interface, \texttt{ChemFit} provides a custom parallelization, using the message passing interface (MPI) to evaluate objective functions, comprised of huge amounts of relatively cheap terms, with low overhead.
In Fig.~\ref{fig:chemfit_code}A, we show the results of a benchmark comparing the performance of various parallelization approaches, implemented using these facilities. Key differences between these approaches are listed in Tab.~S1 of the SI. The benchmark evaluates an objective function with $128$ individual terms (objective function parallelism; see Sec.\ref{subsubsec:obj_func_parallel} for further definitions and details). The total time taken to evaluate the objective function is plotted on the $y$-axis, whereas the time required to evaluate each individual term is shown on the $x$-axis. Our benchmark used $128$ threads or processes, respectively, equal to the number of constituent terms.

We observe that once the evaluation time of individual sample points exceeds approximately 1 ms, both the threadpool (purple) and processpool (green) implementations outperform serial evaluation (blue). Before the crossover point, the serialization and communication overhead of the processpool overshadows the benefits of parallelization. In general it is only necessary to use a processpool when running pure Python code, which is unable to release the global interpreter lock (GIL). On the other hand, Dask (orange) is only beneficial after $2$ ms. Note that the custom MPI (red) is faster than the serial evaluation at all times tested. We remark that $2$ ms is an \emph{extremely} low wall time for many simulation workloads, and wall time ranging from several seconds to many hours are much more common. If individual samples can indeed be evaluated within a millisecond, our recommendation would be to either batch the evaluation or to avoid using \texttt{ChemFit} and, most likely, Python altogether. For longer time, greater than 1 s, all avenues of parallelism explored in this benchmark are expected to be helpful and converge to the same overall wall-time. 

%%%%%%%%%%%%%%%%%%%%%%%%%%%%%%%%%%%%%%%%%%%%%%%%%%%%%%%%%%%%%%%%%
%% Applications
%%%%%%%%%%%%%%%%%%%%%%%%%%%%%%%%%%%%%%%%%%%%%%%%%%%%%%%%%%%%%%%%%

\begin{figure*}[t!]
    \centering
    \includegraphics[width=\linewidth]{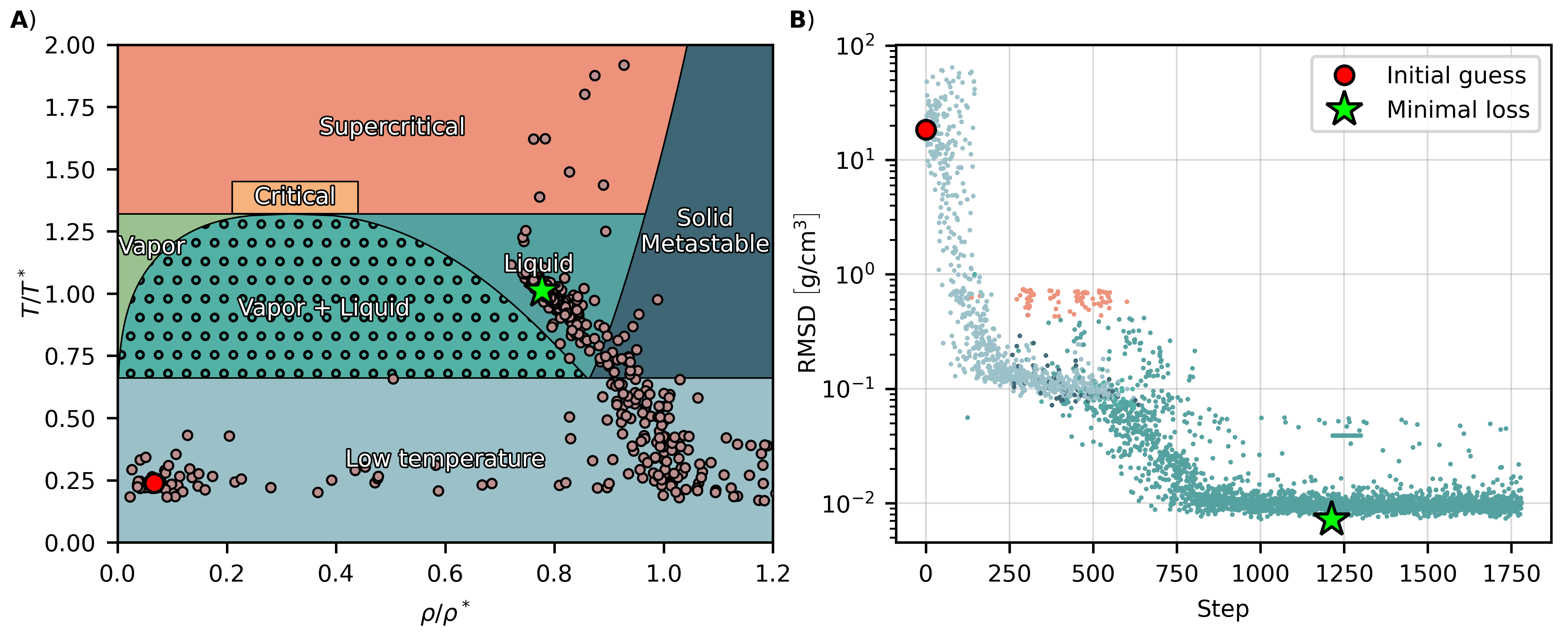}
    \caption{\textbf{A)} ``Trajectory" (brown dots) of the optimization in the space of reduced density ($x$-axis) and reduced temperature ($y$-axis) for  pressure $P=340.23$ atm, temperature $T=120.18$ K and measured density $\rho=1.32$ g/cm$^3$~\cite{streett_1969_experimental}. The data is overlaid upon the phase-diagram of the Lennard-Jones system as per \citet{stephan_2019_thermophysical}.
    The initial guess is marked with a filled red circle and the point with minimal root-mean-square deviation (RMSD) over all 139 experimental points with a green star. \textbf{B)} RMSD versus optimization step. For each step two points were run in tandem, so that the the total number of simulations corresponds to twice the number of steps. The points are colored according to the phase in which \textit{the majority} of the 139 physical systems lie. The color code for each phase is the same as in the left panel.}
    \label{fig:lj_phase_diagram}
\end{figure*}

\subsection{Rediscovering the Lennard-Jones potential parameters for modelling liquid argon}
\label{sec:results_lj}

We first employ \texttt{ChemFit} to revisit the prototypical case of liquid argon described by the Lennard-Jones potential~\cite{lennard1931cohesion}. Since its original introduction by \citet{lennard1931cohesion}, precisely for the description of liquid argon, numerous parameterizations have been proposed~\cite{Rahman1964,Barker1971,Rowley1975,White1999,MendezBermudez2022}. 

Despite its largely idealized nature, few model systems have been studied as extensively as ``Lennard-Jonesium''~\cite{lenhard_2024_childa,Hoef1999,Wang2020,goujon2014gas,sanchez2024predictions,blas2008vapor}. Owing to its conceptual simplicity and broad applicability, the Lennard-Jones potential form has served as the basis for a vast number of interatomic and molecular force fields. Here, we demonstrate how \texttt{ChemFit} can be used to efficiently recover the parameters for liquid argon in a fully automated manner. More importantly, the same workflow readily generalizes to more sophisticated interaction potentials and multiscale models.

We determine the binding energy, $\varepsilon$, and the molecular diameter, $\sigma$, by fitting to experimentally measured densities of liquid argon in the range of 100.9 K to 143.1 K and pressure up to 680 atm, as obtained by \citet{streett_1969_experimental}. In total, there are $139$ independent measurements. We intentionally choose initial parameters which are far away from the known literature values~\cite{Rahman1964,Barker1971,Rowley1975,White1999}, and which do not correspond to the liquid phase~\cite{stephan_2019_thermophysical}. 

\subsubsection{A loss based on the difference between experimental and simulated densities}

For the loss function, we use the root mean square deviation (RMSD) between the simulated densities and experimental densities by \citet{streett_1969_experimental}
\begin{equation}
    O(\varepsilon, \sigma) = \sqrt{\frac{1}{N} \sum_{i=1}^{N} \left(\rho_i(\varepsilon, \sigma) - \rho^\text{exp}_i\right)^2},
\end{equation}
where $N$ is the total number of data points (here, $N=139$), $\rho_i$ and $\rho^\text{exp}_i$ represent the simulated and experimental density, respectively. The temperature and pressure are given by $T_i$ and $P_i$. An overview of the \texttt{ChemFit} workflow used to dispatch individual LAMMPS processes and to evaluate the loss function is shown in Fig.~\ref{fig:chemfit_code}B.

\subsubsection{Explored parameter space and validation}

Fig.~\ref{fig:lj_phase_diagram}A, shows how one of the simulated temperature and pressure pairs moves through the phase-space of the LJ system in the course of the optimization with \texttt{ChemFit}. The optimization starts from the low temperature region of the phase diagram (red circle in the light-blue region in Fig.~\ref{fig:lj_phase_diagram}A), and ends with parameters in the liquid phase region, denoted by a green star in Fig.~\ref{fig:lj_phase_diagram}A. Fig.~S1 in the SI depicts the good agreement of the simulated densities using our optimal parameters with the 139 experimental density points.

During the course of the optimization, the loss reduces by more than three orders of magnitude, as shown in Fig.~\ref{fig:lj_phase_diagram}B, which depicts the loss versus the optimization step. The color scheme corresponds to the phase to which the majority of the simulated systems belongs. Between steps $250$ and $600$, the loss plateaus. In order to escape the plateau, the binding energy, $\varepsilon$, is varied, causing the systems to ricochet between the supercritical, solid metastable and liquid phases, ultimately settling in the liquid phase. In Tab.~\ref{tab:LJ_params} we compare the final results of our optimization process to previously reported $\varepsilon$ and $\sigma$ values (also shown in Fig.~S2 in the SI). Our results are very close to those proposed by \citet{Rahman1964} and \citet{Rowley1975}.

\begin{table}[t]
    \centering
    \caption{Suggested LJ parameters, $T^* = \varepsilon / k_{\mathrm{B}}$ and $\sigma$, for liquid argon from selected references in the literature and this work. The initial parameters were $T^*=503$ K and $\sigma=1.5$ \AA.
    The standard deviation was estimated from the fluctuation of the top 10\% best performing trial parameters.}
    \label{tab:LJ_params}
    \begin{tabular}{lrr}
        \toprule
        Source & $T^*$ [K] & $\sigma$ [\AA]\\
        \midrule
        Rahman (1964)~\cite{Rahman1964} &  120 & 3.4\\
        Barker (1971)~\cite{Barker1971} &  142.95 & 3.361\\
        Rowley (1975)~\cite{Rowley1975} &  119.8 & 3.405\\
        White (1999)~\cite{White1999} & 125.7 & 3.345\\
        \textbf{This work} & 118.7 $\pm$ 0.2 & 3.396 $\pm$ 0.001\\       
        \bottomrule
    \end{tabular}
\end{table}

%%%%%%%%%%%%%%%%%%%%%%%%%%%%%%%%%%%%%%%%%%%%%%%%%%%%%%%%%%%%%%%%%%%%%
%% Polarizable water model 
%%%%%%%%%%%%%%%%%%%%%%%%%%%%%%%%%%%%%%%%%%%%%%%%%%%%%%%%%%%%%%%%%%%%%

\subsection{Parameterizing a polarizable force-field for H$_2$O}
\label{sec:results_scme}

Water is one of the most ubiquitous substances on earth---it is vital for virtually all aspects of life. The quest to understand its properties is a fully fledged scientific subfield in its own right, which is highly relevant for industrial, medical and many other applications ~\cite{ball2008water}. Despite the deceptively simple chemical formula of \ce{H2O}, hydrogen bonding enables water molecules to form a multitude of complex structures~\cite{nilsson2015structural, matsumoto2007topological,tanaka2013importance,goswami2021hybrid}. Further, water exhibits several well-documented anomalies~\cite{shi2020anomalies,russo2018water,espinosa2023possible}, making it exceptionally challenging to model.
Over the past fifty years, water has been described by models with varying degrees of realism. In decreasing order of rigor these include ab-initio theories~\cite{gillan2012assessing}, classical atomistic models~\cite{babin_2013_development,abascal_2005_general}, coarse-grained models~\cite{Molinero2008} and implicit solvent theories~\cite{lenart_2007_effective}.

A possible strategy to parametrize a potential energy function for the interaction of \ce{H2O} molecules is a top-down approach in which the parameters are adjusted until some macroscopic properties of water or ice crystals are reproduced to satisfaction. For example, these properties can include the density, the viscosity and the cohesive energy. In spirit, this approach is similar to our first case study of the LJ potential and liquid argon. A very well-known example of such a potential function that reproduces properties of liquid water is the TIP4P/2005 rigid molecule model~\cite{abascal_2005_general}. A bottom-up approach, instead, aims to parametrize the potential from microscopic considerations, such as local structure or the energy~\cite{babin_2013_development,duboue2017hydration,sanchez2023deep}. Typically, such quantities are not obtained experimentally but from higher-level quantum mechanical calculations. A prominent example of a potential function parametrized in this way is the MB-pol model~\cite{babin_2013_development}.

\begin{figure*}[t!]
    \centering
    \includegraphics[width=\linewidth]{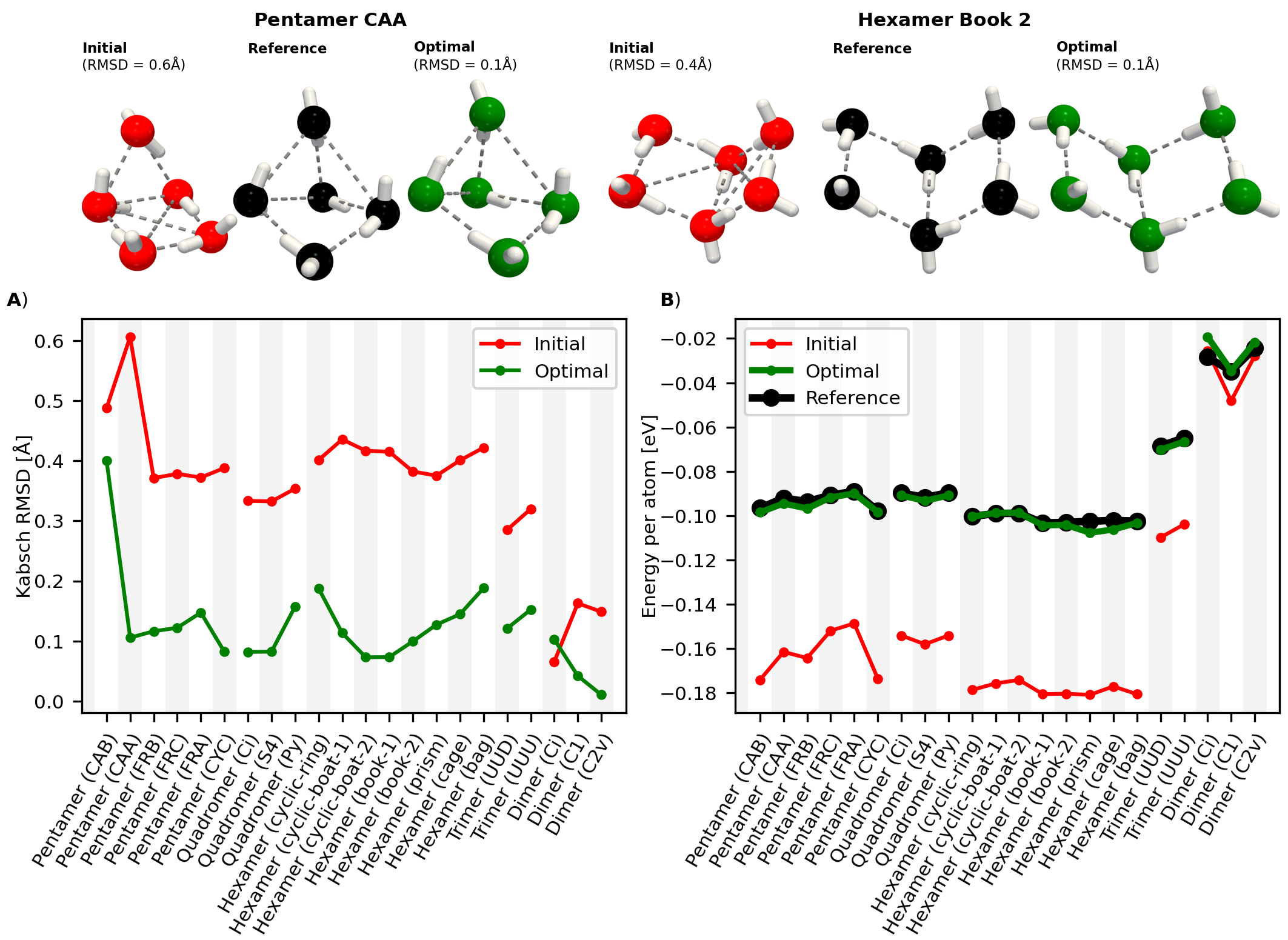}
    \caption{\textbf{A)} Results for the root-mean squared deviation (RMSD) from the Kabsch algorithm~\cite{Kabsch1976, Kabsch1978} for the initial and optimal geometries, compared to reference geometries from density functional theory (DFT) calculations. 
    \textbf{B)} Results for the RMSD from the Kabsch algorithm~\cite{Kabsch1976, Kabsch1978} for the initial and optimal geometries, compared to reference geometries from DFT. 
    \textbf{Inset:} Snapshots of the initial, reference and optimal geometry of the water cage pentamer CAA (left) and the water hexamer book-2 (right).
    Dashed grey lines are a guide for the eye, connecting oxygen atoms that are less than $3.5$ \AA ~apart.}
    \label{fig:scme}
\end{figure*}

Here, we use a bottom-up approach to parametrize a model situated on the more rigorous end of the classical atomistic models: the flexible single-center-multipole-expansion (SCME) potential function for \ce{H2O}~\cite{wikfeldt_2013_transferablea,jonsson_2019_polarizable,dohn_2019_polarizable,jonsson_2022_transferable,myneni2022polarizable}. We design an optimization procedure to find parameters which reproduce the geometry of 22 small reference \ce{H2O} clusters (see, for example, the insets of Fig.~\ref{fig:scme}), whose structures were obtained by locally minimizing the energy as computed by DFT,
using the BEEF-vdW functional~\cite{lundgaard_2016_mbeefvdw} (see Sec.~\ref{subsubsec:scme_methods} for further technical details on these calculations). These clusters are composed of different isomers, ranging from dimers to hexamers. In particular, cyclic water hexamers have been found to be nearly iso-energetic~\cite{Kim1998}, making their geometry an interesting target property for model parameterization~\cite{Petty2025}. We choose the geometry of locally optimal \ce{H2O} clusters as targets, since local water structure strongly influences the behaviour of aqueous electrolytes~\cite{marcus2009effect,duboue2017hydration,goswami2024evidence,weingaertner1984unusual,gomez2022hydrated,Goswami2026}, which are central to many chemical, biological, and industrial processes.
While this example only deals with pure water, it could easily be extended to target the structure of water molecules around a solvated ion.

\subsubsection{Loss function based on optimal structural alignments}
\label{subsubsec:setup_scme}

For a given trial parameter set, our objective function first minimizes the energy of the SCME potential function, with the tested parameters, and then quantifies the deviation from the reference configuration via the RMSD of atomic positions \emph{after} applying an optimal rotation and translation determined with the Kabsch algorithm~\cite{Kabsch1976,Kabsch1978}. A one-to-one correspondence between atoms in the reference configuration and in the minimized configuration is assumed. The overall loss function is simply the sum of the RMSD values for each reference configuration. In order to showcase the optimization process, we optimize the eight parameters listed in Tab.~S2 of the SI, and start from initial values that are different from the reported literature values by \citet{jonsson_2022_transferable}.

\subsubsection{Validation against the density-functional theory energy}
\label{subsubsec:results_scme}

In Fig.~\ref{fig:scme}A, we show the values of the RMSD using the initial (solid red line) and optimal parameters (solid green line). Lower values indicate better agreement with the reference geometry. Essentially, this is a comparison of the contribution of each cluster to the overall loss value, before and after optimization with \texttt{ChemFit}. To contextualize the following RMSD values, we use the equilibrium distance between an oxygen and a hydrogen atom in \ce{H2O}, which is approximately $r_\text{OH} = 0.957$ \AA ---visually, $r_\text{OH}$ corresponds to the length of the white ``sticks" in the insets of Fig.~\ref{fig:scme}. We find that the agreement of the structures is greatly improved after \texttt{ChemFit} optimization---on average by around $20$\% of the $r_\text{OH}$ distance (e.g., $<$0.2 \AA). Only the Dimer Ci structure has slightly worse agreement with the optimal parameters, although the RMSD is still less than $10$\% of the $r_\text{OH}$ distance. The RMSD of several other structures is reduced by $0.3$ to $0.4\, r_\text{OH}$. The optimization also improves the relative orientations of the OH bonds in the structures obtained with the optimized parameters---mostly all OH bonds point in the same direction as those in the reference structures from DFT, as shown in the insets of Fig.~\ref{fig:scme}.

After assessing the geometric agreement between the reference and optimized water clusters, we further validate the optimization by comparing the energy computed using DFT and the SCME model, shown in Fig.~\ref{fig:scme}B. Notably, the energy was \emph{not} included explicitly among the optimization targets. Although some correlation between stationary structures and their energy is naturally expected, agreement in the locations of the minima of two functions---here, the DFT and SCME energy surfaces---does not imply quantitative agreement between the functions themselves. Consequently, this comparison provides a meaningful additional validation metric. As shown in Fig.~\ref{fig:scme}B, the SCME energy obtained from the optimized parameter set (green line) reproduces the reference DFT energy of the water clusters (black line) within \(0.01\) eV per atom. This difference corresponds to less than \(5\%\) of a typical hydrogen-bond energy ($\sim0.216$ eV)~\cite{wendler_2010_estimating}.

%%%%%%%%%%%%%%%%%%%%%%%%%%%%%%%%%%%%%%%%%%%%%%%%%%%%%%%%%%%%%%%%%%%%%
%% CG protein model model
%%%%%%%%%%%%%%%%%%%%%%%%%%%%%%%%%%%%%%%%%%%%%%%%%%%%%%%%%%%%%%%%%%%%%

\subsection{Tuning the critical temperature of a high-resolution coarse-grained protein model}
\label{sec:results_mpipi}

\begin{figure*}[!th]
    \includegraphics[width=\linewidth]{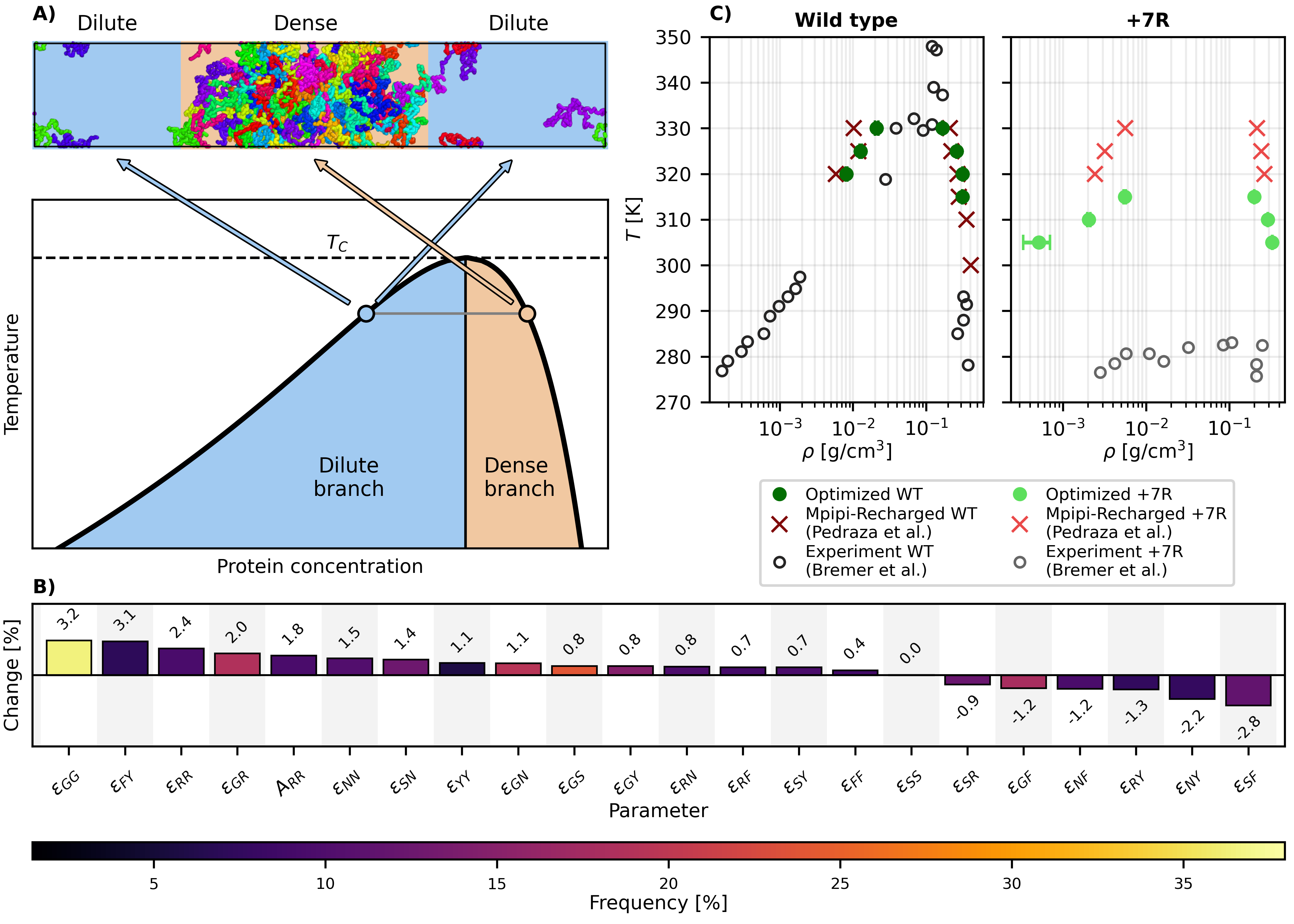}
    \caption{\textbf{A)} \textit{Bottom:} Idealized illustration of a typical phase-diagram showing the coexistence between a dense and a dilute branch. The coexistence curve is plotted as a thick black line, with the critical temperature $T_C$ demarcated by a thin dashed line. Above the critical temperature no coexistence can be observed. The orange(blue) shaded regions are visual guides to highlight the dense(dilute) branch.
    \textit{Top:} Snapshot from a direct-coexistence (DC) simulation, which can be used to obtain coexistence curves. A dense protein condensate coexists with a dilute phase of solvated proteins --- the chemical potential of both phases is equal. The background shades, again, highlight the dense and dilute regions of the simulation box. Arrows indicate how these regions of the simulation box relate to the coexistence curve.
    \textbf{B)} Percentage change relative to the initial Mpipi-Recharged parameters of all parameters which were optimized with ChemFit. The color code corresponds to the geometric mean of all participating residues over the WT and +7R variant.
    \textbf{C)} Coexistence points in the density-temperature plane for the wild-type(+7R variant) of the complexity domain (LCD) of hnRNPA1 are shown on the left(right).
    Black(grey) circles indicate experimental data of the WT(+7R) protein by \citet{bremerDecipheringHowNaturally2022}, while the Mpipi-Recharged simulations by \citet{pedrazaPredictingSaturationConcentrations2025a} are shown by maroon(pink) crosses. The results for the WT(+7R) protein, after tuning the critical temperature $T_C$ with \texttt{ChemFit}, are indicated by dark green(light green) filled circles.
    }
    \label{fig:mpipi}
\end{figure*}

We now turn to a biophysical application: the phase separation behaviour of protein condensates. Recent studies have shown that the dynamic formation of biomolecular condensates is an essential aspect of biological function and cell compartmentalization~\cite{brangwynne_2009_germline,hyman2014liquid,Shin2017_science,BOEYNAEMS2018420}. However, its misregulation can result in aberrant progressive  condensate solidification, which has been linked to several neurodegenerative disorders, such as Parkinson's disease, Huntington's chorea and Alzheimer's disease~\cite{Alberti2021,visser2024role}.
Therefore, understanding the coexistence phase behaviour of protein condensates and their stability and material properties is an interdisciplinary research area of intense activity~\cite{mitrea2022modulating}.

An archetypal protein consists of a sequence of amino acids, which are of 20 different natural types. This chain-like character makes proteins similar to polymers.
Numerous proteins contain intrinsically disordered regions (IDRs), without a fixed secondary or tertiary structure---colloquially speaking, they are ``floppy".
In multi-domain proteins, the IDRs are interspersed with globular regions, possessing secondary and tertiary structure, such as $\alpha$-helices and $\beta$-sheets. If a protein lacks these rigid regions, it is called an intrinsically disordered protein (IDP). It is typically the multi-valency of interactions between IDRs or IDPs that drives phase separation~\cite{martin2020intrinsically,tsang2020phase,sipko2024multivalency,feito2024capturing}. In this case, phase separation is characterized by the coexistence of a dense phase, which is the protein condensate, and a dilute phase of solvated proteins (displayed in the inset of Fig.~\ref{fig:mpipi}A). An idealized picture of such a coexistence curve is depicted in the phase-diagram in Fig.~\ref{fig:mpipi}A. The upper critical temperature, $T_C$, is the temperature above which there is no coexistence, and hence no separation into distinct phases (dashed line in Fig~\ref{fig:mpipi}A). Thus, by knowing $T_C$ alone, one can determine whether phase separation occurs at physiological conditions, or not. 

In this case study, we focus on the low-complexity domain (LCD) of hnRNPA1, an RNA-binding protein abundant in motor neurons that has been linked to amyotrophic lateral sclerosis (ALS)~\cite{bampton2020role,lee2024mutation}. We optimize a subset of parameters of the residue-resolution coarse-grained (CG) Mpipi-Recharged model~\cite{R.Tejedor2025} (model details in Sec.~\ref{subsubsec:mpipi_model}).
A recent study by \citet{pedrazaPredictingSaturationConcentrations2025a} demonstrated near-quantitative agreement between predictions of this model and \textit{in vitro} experiments by \citet{bremerDecipheringHowNaturally2022} for the hnRNPA1 WT LCD and several engineered mutant variants, all of which are intrinsically disordered proteins (IDPs). For mutant variants not enriched in arginine, the average discrepancy between simulated and experimental $T_C$ values was below $7$ K, whereas for the two arginine-enriched variants, +7R and +7R+12D, the discrepancy increased to approximately $56$ K (see Tab.~S3 in the SI).
The goal of this case study is therefore to improve the predicted $T_C$ of the +7R variant while preserving the excellent agreement obtained for the WT sequence~\cite{bremerDecipheringHowNaturally2022,alshareedahSequencespecificInteractionsDetermine2024}.

\subsubsection{Loss function based on density variations around the critical temperature}
\label{subsubsec:setup_mpipi}

We optimize $22$ out of the approximately $800$ model parameters of the Mpipi-Recharged. The optimized parameters include the binding energy, $\varepsilon_{ij}$, and Yukawa prefactor, $A_{ij}$ of cross and self interactions between arginine and the most abundant residues in the hnRNPA1 WT and +7R variant sequences (see Sec.~\ref{subsubsec:mpipi_model} for further details on the model parameterization). To prevent overfitting, we impose bounds of $15\%$ around the original value of each parameter.

Direct-coexistence (DC) simulations are an established method for determining whether a protein condensate is stable or breaks apart ~\cite{ladd1977triple,espinosa2013fluid,espejo2025compositional}. Reminiscent of experiments, in DC, a dense phase is placed within an elongated simulation box and allowed to equilibrate, eventually partitioning into a condensate and dilute phase, or becoming a single phase when the temperature is above the critical one. However, such simulations entail the use of a large enough simulation box to prevent finite size effects, and are also usually run for relatively long time (typically $2-3$ $\mu$s) to sample the equilibrium concentration of the dilute phase. Therefore, DC simulations are an impractical choice for the objective function in \texttt{ChemFit}.

The chosen objective function instead acts as a cheaper proxy for DC. It is designed to maximize the jump in density, $\rho$, occurring when the experimentally expected critical temperature, $T^\text{exp}_C$, is crossed in an isotherm at a pressure of $P=0$ bar. Typically, NPT simulations at $P \approx 0$ and $T < T_C$ correspond to the dense branch of the coexistence curve~\cite{pedrazaPredictingSaturationConcentrations2025a,sanchez2025charged}. For each sequence, we define an objective $O_\text{seq}$ as

\begin{equation}
    O_\text{seq} = -\frac{\rho_\text{seq}(T^\text{seq}_\text{lo}) - \rho_\text{seq}(T^\text{seq}_\text{hi})}{\rho_\text{seq}(T^\text{seq}_\text{lo}) + \rho_\text{seq}(T^\text{seq}_\text{hi}) },
    \label{eq:obj_tc}
\end{equation}
with $\text{seq} \in [\text{WT},\text{+7R}]$, resulting in an overall objective function of
\begin{equation}
    O = O_\text{WT} + O_\text{+7R}.
    \label{eq:obj_tc_total}
\end{equation}

The two temperatures $T^\text{seq}_\text{lo}$ and $T^\text{seq}_\text{hi}$ are chosen such that they are a certain $\Delta T$ below and above the \emph{experimental} critical temperature of that sequence, leading to 
\begin{equation*}
    T^\text{seq}_\text{lo} = \left[ T^\text{seq}_C\right]_\text{exp} - \Delta T
\end{equation*}
and
\begin{equation*}
     T^\text{seq}_\text{hi} = \left[ T^\text{seq}_C\right]_\text{exp} + \Delta T.
\end{equation*}

For these calculations we use $\Delta T=10$ K. A more detailed justification and discussion of this objective function is given in Sec.~IVB of the SI.

\subsubsection{Model optimization and validation}
\label{subsubsec:results_mpipi}

We show the final obtained Mpipi parameter values in Tab.~S4 of the SI and a visual overview of the relative change of each chosen parameter, which is within about $5\%$, in Fig.~\ref{fig:mpipi}B. The color scheme corresponds to the geometric mean of all optimized self and cross interactions in the model, indicating their relative abundance across the WT and +7R variants. Interestingly, the relative abundance of participating residues is not always directly correlated with the relative change in the optimized parameters with respect to the original Mpipi-Recharged force field. For instance, while both sequences are rich in glycine and $\varepsilon_{GG}$ exhibits a large change, $\varepsilon_{FY}$ also changes by almost the same amount but corresponds to a low abundance of around $5\%$ across the sequence.

In order to validate the results, we performed DC simulations with the optimized parameters. In Fig.~\ref{fig:mpipi}C we compare the phase diagram obtained through DC results with simulations with the original Mpipi-Recharged parameters (obtained by \citet{pedrazaPredictingSaturationConcentrations2025a}) and with the experimental measurements by \citet{bremerDecipheringHowNaturally2022}.
The canonical Mpipi-Recharged model (maroon crosses) predicted close agreement with experimental values from \citet{bremerDecipheringHowNaturally2022} (black open circles) for the WT protein, within the uncertainty of DC simulations. This agreement is maintained with the optimized parameters (dark green circles). For the +7R variant, the divergence between the Mpipi-Recharged model (red crosses) and experiments (gray open circles) is more drastic, about $56$ K. 

The optimized model (light green circles) fares better, predicting a $T_C$ that is about $25$ K lower than the Mpipi-Recharged prediction. The new model $T_C$ still diverges by about $30$ K, but it qualitatively recovers the experimental trend more closely without compromising the model performance for the WT sequence. We note that achieving quantitative agreement for both variants would require optimizing the full set of approximately 800 parameters in the model, as well as evaluating its performance across more than 100 protein sequences, as carried out in the original Mpipi-Recharged study~\cite{R.Tejedor2025}. Such an undertaking, however, lies well beyond the scope of the present work, whose aim is instead to demonstrate the applicability of \texttt{ChemFit} as a proof-of-concept for biomolecular coarse-grained model optimization for condensate phase behaviour.

%%%%%%%%%%%%%%%%%%%%%%%%%%%%%%%%%%%%%%%%%%%%%%%%%%%%%%%%%%%%%%%%%%%%%
%% Conclusions
%%%%%%%%%%%%%%%%%%%%%%%%%%%%%%%%%%%%%%%%%%%%%%%%%%%%%%%%%%%%%%%%%%%%%
\section*{Discussion}

We have presented \texttt{ChemFit}, a flexible and scalable framework for defining and evaluating simulation-based objective functions in parameter optimization problems. \texttt{ChemFit} enables the effective application of gradient-free and black-box optimization methods to problems characterized by expensive, noisy, and heterogeneous simulations. We show how \texttt{ChemFit} is applicable across a broad range of scientific domains and simulation scales.

First, our LJ case study illustrates how optimal parameters for the argon phase diagram can be found even when starting from a very different region of the parameter space with a noisy objective function (based on the solution density), showing how \texttt{ChemFit} can be helpful in parameterizing novel systems. Although the densities are sufficient for our example, additional target properties of interest, such as solubility, surface tension or viscosity~\cite{Zeron2019,MendezBermudez2022} can be combined in \texttt{ChemFit}.

In our second example, we adjusted the parameters of a realistic atomistic flexible and polarizable model for the interaction between \ce{H2O} molecules. Similar tasks frequently arise in the study of small molecules in water solution, for example, aqueous electrolytes~\cite{andreevInfluenceIonSolvation2018,panagiotopoulosDynamicsAqueousElectrolyte2023,Goswami2026,sanchez2023direct}. Quantifying all the possible cross-interactions is often necessary, because Lorentz-Berthelot mixing rules frequently do not deliver quantitatively satisfying results~\cite{Zeron2019}. But, even for solutions with only a few components, the number of adjustable parameters can be vast---making an automated approach extremely valuable in terms of model accuracy and optimization time. Using structures obtained from explicit 
DFT simulations as a reference is interesting because this enables explicitly targeting local structure of ice and water, as opposed to parameterizing against bulk quantities, such as the density (as in our first example). Local structure can be particularly important for modelling accurate solvation behaviour ~\cite{andreevInfluenceIonSolvation2018,weingaertner1984unusual,goswami2024evidence,Goswami2026}. Our parameterization approach is also applicable for hybrid methods, such as QM/MM~\cite{jonsson_2022_transferable,dohn_2019_polarizable}, wherein interactions between the MM and QM region need to be finely tuned. The direct integration of \texttt{ChemFit} with ASE, as highlighted in the SCME example, is useful because it opens up immediate access to several DFT codes. More specifically, the SCME parameterization example shows how structure optimization against high-level calculations can be used to obtain accurate agreement with the energy, indirectly proving the transferability of the SCME potential. Our optimized parameters (our reference data used the BEEF-vdW functional) are consistent with independent parameters derived by \citet{jonsson_2022_transferable}, which were based on quantum mechanical coupled-cluster reference data \cite{bartlett_2007_coupledcluster}. 

Lastly, we refined the predicted critical temperature of the Mpipi-Recharged model \cite{R.Tejedor2025} for two variants of the LCD of hnRNPA1. The critical temperature is an emergent property of protein solutions that requires several simulations to compute and is inherently noisy and difficult to be quantitatively reproduced~\cite{feito2025benchmarking,pedrazaPredictingSaturationConcentrations2025a}. The automation of large-scale parameter space exploration, as enabled by \texttt{ChemFit}, yielded results which are nearly intractable to obtain with manual approaches. We note that even seemingly small changes as performed during this model optimization can lead to surprisingly large changes in macroscopic condensate phase behaviour. This highlights the difficulty of tuning models in high-dimensional parameter spaces. Our approach could also be of relevance to the parameterization of cross-interactions between protein, RNA~\cite{perez_2004_relative,Tejedor2026RNA} and DNA chromatin models~\cite{farr2021nucleosome,russell2025near,ivani_2016_parmbsc1}, as well as explicit ions in  coarse-grained protein models~\cite{garaizar2021salt}.

Certain CG protein models are parameterized against single-protein properties, such as the radius of gyration~\cite{tesei2021accurate}. However, faithfully reproducing the properties of isolated proteins may not always translate to accurate predictions of collective macroscopic behaviour, such as phase separation~\cite{feito2025benchmarking}. Therefore, we believe that it is valuable to parameterize CG protein models against such emergent behaviour. Of course, the radius of gyration, end-to-end distance, $\theta$ temperature and other properties can also be incorporated into the optimization using \texttt{ChemFit}. The $\theta$ temperature, specifically, could be an interesting single-protein target property, due to its strong correlation with the critical solution temperature~\cite{dignon2018relation,garaizar2020expansion}. 

Overall, the presented examples convincingly demonstrate the broad applicability of \texttt{ChemFit} in applications that combine high-dimensional parameter spaces with increasingly complex and computationally demanding models. We believe that the \texttt{ChemFit}  platform not only reduces human effort in model development but also presents new opportunities for combining models from different scientific domains, with minimal friction and maximum accuracy.

%%%%%%%%%%%%%%%%%%%%%%%%%%%%%%%%%%%%%%%%%%%%%%%%%%%%%%%%%%%%%%%%%%%%%
%% Methods section
%%%%%%%%%%%%%%%%%%%%%%%%%%%%%%%%%%%%%%%%%%%%%%%%%%%%%%%%%%%%%%%%%%%%%

\section{Methods}

\subsection{Software Design}

Detailed explanations of how to use \texttt{ChemFit} can be found in the online documentation~\cite{chemfit_docs} and examples~\cite{chemfit_examples}.

\subsubsection{Objective functions}

For the purpose of \texttt{ChemFit}, an objective function is any kind of function that maps a set of input parameters to a single scalar value --- the loss. The purpose of optimization algorithms is to find parameters, which minimize this loss value.

\texttt{ChemFit} splits the computation of the loss value into two steps: 
\begin{enumerate}
    \item The computation of intermediate \emph{quantities} via explicit simulations. 
    For instance these may be the energy, density or any observable obtained from computationally expensive (molecular) simulations, while the \emph{parameters} could be the model parameters of the force-field being parameterized.
    \item The \emph{loss} is computed by applying a function that maps from the quantities (and the parameters) to a single scalar value.
    As a rule of thumb, these quantities are expensive to compute, while the loss is cheap (if the quantities are known). 
\end{enumerate} 

We can summarize this procedure as follows
\begin{equation*}
    \text{Parameters} \underbrace{\longrightarrow}_{\text{expensive}} \text{Quantities} \underbrace{\longrightarrow}_\text{cheap} \text{Loss Value}.
\end{equation*}

This split decouples the execution of computationally expensive simulations from the exact quantity which drives the optimization, and enables the accumulation of by-products of the simulation (in meta-data) which do not directly influence the optimization. Further, it increases flexibility since, once the mapping of parameters to quantities has been established, the loss function can be interchanged easily.

\texttt{ChemFit} encodes this relationship in an abstract interface, called a \emph{QuantityComputer}.
Out of the box, the framework comes with three pre-defined \emph{QuantityComputer}s:

\begin{enumerate}
    \item The \emph{FileBasedQuantityComputer}, which can run arbitrary executables and coordinates user-supplied parsers to generate the quantities.
    \item The \emph{SinglePointASEComputer}, which runs any ``Calculator" defined within the atomic simulation environment (ASE) \cite{hjorthlarsen_2017_atomic} on a reference atom configuration. 
    \item The \emph{MinimizationASEComputer}, which relaxes a configuration, using ASE, to a local minimum before evaluating the quantities.
\end{enumerate}
    
Furthermore, additionally arbitrary user-defined \emph{QuantityComputer}s can be easily added. Once a \emph{QuantityComputer} is defined, it needs to be combined with a loss function, e.g. a simple root mean square deviation (RMSD), and the objective function is ready to use. 

\subsubsection{Interface to optimization libraries}
While the primary purpose of \texttt{ChemFit} is the definition of objective functions, which can be used with any optimization algorithm available through Python, we also provide a \textit{Fitter} class. It enables access to Scipy~\cite{2020SciPy-NMeth} and Nevergrad~\cite{nevergrad} through a unified interface and provides a few useful features such as a callback system, robustness features against exceptions, a checkpointing and a yrestart functionality.

\subsection{Concurrency}
\label{subsec:concurrency}

Concurrency enters the optimization procedure in three different ways, described in the following sections: (i) Simulation engine parallelism, (ii) Objective function parallelism and (iii) Parameter trial parallelism.

Computational resources should be allocated to these modes in the order listed above, only moving to the next mode once the limitations of the current mode are reached. In other words, individual simulations should be parallelized up to the strong scaling limit, after which resources should then be used to parallelize the computation of the objective function of sample points, and so on.

\subsubsection{Simulation engine parallelism} \label{subsubsec:sim_parallel}

At the most basic level, most simulation codes, such as LAMMPS~\cite{Thompson2022}, GROMACS~\cite{Abraham2015} or VASP~\cite{kresse_1996_efficient}, are themselves able to utilize multiple threads and/or processes. If the ``size" of the system is held constant, however, there is a soft upper limit up to which computational resources without suffering from significant diminishing returns---this limit is dictated by the strong scaling efficiency of the code in question~\cite{Spiteri2025,Raffenetti2017}.

\subsubsection{Objective function parallelism} \label{subsubsec:obj_func_parallel}

If the objective function entails computations for multiple sample points---for instance, say the objective function is a combined function of 100 points---the simulations for these points can be run in parallel (for a particular parameter trial set). In \texttt{ChemFit} this is achieved by defining a \emph{CombinedObjectiveFunction} and parallelizing it either via the message passing interface (MPI) or a Python executor, such as the ``processpool" and ``threadpool" provided by the standard library, or third party libraries like ``dask"~\cite{rocklin2015dask}. While objective function concurrency has excellent strong scaling efficiency, since all samples are independent, it is limited in two ways: (i) the wall time can never be lower than the slowest sample point, which for extremely long simulations can severely limit the number of trial parameter sets tested, and (ii) the parallelism is limited by the number of sample points.

\subsubsection{Parameter trial parallelism} \label{subsubsec:parameter_parallel}

Lastly, the objective function can be evaluated for multiple candidate parameter sets in parallel. This mode of concurrency has excellent strong scaling efficiency and, in contrast to simulation engine parallelism (Sec.~\ref{subsec:concurrency}) and objective function parallelism (Sec.~\ref{subsubsec:obj_func_parallel}), there is no practical limit to the amount of parallelism which can be leveraged in this way. 

However, a significant drawback can be illustrated by considering that two trials in sequence are more valuable than two trials in parallel. This is because the second trial can use the result of the first trial to refine its guess---the generalization to multiple trials is obvious. On one hand, exploiting this mode of concurrency is a detail of the optimization algorithm---most black-box algorithms support it trivially via some variation of an ``ask-and-tell" interface. On the other hand, objective functions that support correct parallel evaluations need to avoid so-called race conditions, occurring when multiple threads or processes access shared resources simultaneously leading to undefined behaviour.
To avoid race conditions, \texttt{ChemFit} provides a unique \emph{EvaluateContext} per objective function evaluation. This means one objective function can be evaluated in $N$ threads or processes by instantiating $N$ \emph{EvaluateContext}s, which encapsulate the access to shared resources.

\subsection{Molecular-dynamics workflow for fitting Lennard-Jones parameters to argon densities}

The Lennard-Jones (LJ) potential~\cite{lennard1931cohesion} is described by:
\begin{equation}
    U_\text{LJ}(r) = 4\varepsilon \left[ \left(\frac{\sigma}{r}\right)^{12} - \left(\frac{\sigma}{r}\right)^6 \right],
\end{equation}
where $r$ is the distance between two interacting argon atoms, $\sigma$ corresponds to the size of the atoms, and $\varepsilon$ represents the interatomic binding energy. The attractive $1/r^6$ term corresponds to the London dispersion forces~\cite{london1930theorie, Jones1924}, derived from quantum mechanical perturbation theory and the repulsive term is phenomenological~\cite{Kulakova2017}. 
% Even though the success of the LJ potential in describing liquid argon is likely due to a lucky coincidence~\cite{Hoef1999, Wang2020,goujon2014gas}, its popularity and extensive application to other systems make it a natural choice for this example.

As depicted in Fig.~\ref{fig:chemfit_code}B, a \emph{CombinedObjectiveFunction} is used to tie together individual terms expressed via the \emph{FileBasedQuantityComputer} class, which handles the dispatch of individual LAMMPS processes and the parsing of the output files. 

We use LAMMPS~\cite{Thompson2022} to generate the molecular dynamics (MD) trajectories. Given the fact that the optimization with \texttt{ChemFit} needs to evaluate a wide range of different $\varepsilon$ and $\sigma$ candidates, it is important to have a well-crafted and robust LAMMPS script. Details of the simulation setup for the LJ calculations can be found in Sec.~IIA of the SI.

We ran \texttt{ChemFit} on a single cluster node with 128 cores. Per evaluation of a parameter set ($\varepsilon$, $\sigma$), which involves the computation of 139 densities, we used 64 cores. To evaluate each density for a given ($\varepsilon$, $\sigma$) we ran LAMMPS with a single core.
Thus, we evaluated two parameter sets concurrently at all times---saturating the entire node, thereby leveraging objective function parallelism (see also Sec.~\ref{subsubsec:obj_func_parallel}). 
We evaluated 1780 steps for a total of 3560 trial parameters using Nevergrad's ``NgIohTuned" optimizer~\cite{nevergrad}. The wall time was 62 hours.

\subsection{Geometry-based optimization of the SCME/f polarizable H$_2$O model}

\subsubsection{The SCME potential}
\label{subsubsec:scme_methods}

In the SCME potential, the electrostatics of a water molecule are described by a static Cartesian multipole expansion ranging from dipole up to, and including, hexadecapole. By comparison with electronic structure calculations of ice and water clusters, a truncation at the hexadecapole was found to be adequate~\cite{batistaElectricFieldsIce2000}.
Additionally, dipole-dipole, dipole-quadrupole and quadrupole-quadrupole polarizabilities are included giving rise to induced dipole and quadrupole moments \cite{wikfeldt_2013_transferablea}. 
We used the flexible variant of the SCME potential, called SCME/f, introduced by \citet{jonsson_2019_polarizable}, which further implements a dependence of the static dipole and quadrupole moment on the geometry of the \ce{H2O} molecule (OH bond lengths and angle) balanced by an energy term favoring the equilibrium geometry described by the Partridge-Schwenke~\cite{partridge_1997_determination} potential energy surface of the water monomer.
The damping of the multipole interactions at short range is described by a Gaussian smearing function \cite{stone_2011_electrostatica},
The damping length is controlled by a free parameter $\tau_E$, which among others, we will optimize with \texttt{ChemFit}. A more detailed explanation of the electrostatic damping can be found in Sec.~IIIA of the SI. 

In addition to the electrostatic interactions, the SCME/f force field includes the dispersion interaction
\begin{equation}
    \begin{split}
        E_\text{Disp} = &\frac{1}{2}\sum_{ij} \left[ \frac{C_6}{r_{ij}^6}g_6(r_{ij}) \right.\\
        + &\left.\frac{C_8}{r_{ij}^8}g_8(r_{ij}) + \frac{C_{10}}{r_{ij}^{10}}g_{10}(r_{ij})\right],
    \end{split}
\end{equation}
where the indices $i$ and $j$ label the oxygen positions, $r_{ij}$ is the distance between $i$ and $j$; $C_6$, $C_8$ and $C_{10}$ are constants quantifying the correlation between instantaneously induced moments due to fluctuating densities including e.g. 
dipole-dipole, dipole-quadrupole, dipole-octupole and quadrupole-quadrupole polarizabilities. 
Moreover, $g_{n}(r)$ is a short ranged damping function, following the Tang-Toennis form \cite{tang_1984_improved}
\begin{equation}
    g_n(r) = 1 - e^{-\tau_d r} \sum_{k=0}^n \frac{(\tau_d r)^k}{k!},
\end{equation}
where $\tau_d$ is another free parameter---the damping strength. In the original SCME/f, the coefficients $C_6$ to $C_{10}$ were adopted from~\citet{wormer1992many}.
Lastly, the repulsive core contribution is given by a modified Born-Mayer potential~\cite{jonsson_2022_transferable}
\begin{equation}
    E_\text{Rep} = \frac{1}{2}\sum_{ij} A \left(\frac{r_{ij}}{L}\right)^{B} e^{C r_{ij}},
    \label{eq:scme_repulsion}
\end{equation}
where $A$, $B$, $C$ are free parameters and $L = 1$ \AA~is a constant introduced to simplify unit conversions.
The full details of the formulation of the electrostatics in the SCME/f potential are best studied from dedicated references~\cite{wikfeldt_2013_transferablea,jonsson_2022_transferable,myneni2022polarizable}.

\subsubsection{Setup details}

Since the SCME/f potential is available as a ``Calculator" in the atomic simulation environment (ASE)~\cite{hjorthlarsen_2017_atomic}, we used the built-in \emph{MinimizationASE} quantity computer, which minimizes the energy of an initial reference configuration and then computes certain quantities.

For the reference configurations, we used the locally optimal configurations obtained from structure optimization from DFT calculations with the BEEF-vdW functional \cite{lundgaard_2016_mbeefvdw}, as implemented in VASP~\cite{kresse_1996_efficient}. 

We ran the optimization on a single computer using four cores. The optmizer was Nevergrad's~\cite{nevergrad} ``NgIohTuned" optimizer for 100 steps.
Since this objective function contains no noise, this was followed by a final relaxation with the Limited-Memory-Broyden-Fletcher-Goldfarb-Shanno (LBFGS) algorithm as implemented in Scipy~\citet{2020SciPy-NMeth}. The wall time was 30 minutes.

\subsection{Direct-coexistence simulations for tuning Mpipi-Recharged critical temperatures}

\subsubsection{The Mpipi-Recharged model}
\label{subsubsec:mpipi_model}

The Mpipi-Recharged model belongs to a class of transferable implicit-solvent residue-resolution CG protein models~\cite{Joseph2021,tesei2021accurate,tesei2023improved,Cao2024, das2020comparative,dignon2018sequence,desancho_2007_evaluation}, wherein amino acids are represented by a single spherical bead. Residue-pair interactions are defined by combinations of short-range and long-range potentials, describing hydrophobic and ion-screened Coulomb interactions, respectively~\cite{R.Tejedor2025}. Water and solvated ions are modelled as an implicit solvent. 
% moved to methods, shortened here.
% The Mpipi-Recharged model, in particular, has been demonstrated to provide highly competitive predictions of single-protein properties such as the radius of gyration, and biomolecular phase separation, including condensate stability or relative variations in material properties as a function of sequence variations ~\cite{pedrazaPredictingSaturationConcentrations2025a,R.Tejedor2025,w7g3-6rsd,espejo2025compositional}. 
% The Mpipi-Recharged model has been demonstrated to provide highly-competitive predictions of single-protein properties and biomolecular phase separation~\cite{pedrazaPredictingSaturationConcentrations2025a,R.Tejedor2025,w7g3-6rsd,espejo2025compositional}.

The Mpipi-Recharged model has been shown to provide near-quantitative predictions of single-protein properties such as the radius of gyration, and biomolecular phase separation, including condensate stability or relative variations in material properties as a function of sequence variations ~\cite{pedrazaPredictingSaturationConcentrations2025a,R.Tejedor2025,w7g3-6rsd,espejo2025compositional}. 
In the Mpipi-Recharged model, short-range hydrophobic interactions are represented by the Wang-Frenkel potential~\cite{Wang2020}:

\begin{equation}
    \begin{split}
        E_\text{WF}(r_{ij}) = \varepsilon_{ij}\alpha_{ij}
        \!\!
        &\left[ 
            \left(
                \frac{\sigma_{ij}}{r_{ij}}
            \right)^{\!\!2\mu_{ij}} - 1
        \right]\\
        \times
        &\left[ 
            \left(
                \frac{r_c}{r_{ij}}
            \right)^{\!\!2\mu_{ij}} - 1
        \right]^{2\nu_{ij}}\kern-1em,
    \end{split}
\end{equation}
with 
\begin{equation*}
    \alpha_{ij} = 2\nu_{ij} \left(\frac{r_c}{\sigma_{ij}}\right)^{\!\! 2\mu_{ij}}
    \left[
        \frac{1+2\nu_{ij}}{2\nu_{ij}[(r_c / \sigma_{ij})^{2\mu_{ij}} - 1]}
    \right]^{2\nu_{ij} + 1}\kern-2em,
\end{equation*}
where $\varepsilon_{ij}$ is the binding energy, $r_c$ is the cutoff radius, $\sigma_{ij}$ is the bead size parameter and $\nu_{ij}$ and $\mu_{ij}$ are (integer) constants.

The screened electrostatics, on the other hand are specified by a Yukawa potential between charged residues
\begin{equation}
    E_\text{Y}(r_{ij}) = \frac{A_{ij}}{r_{ij}} e^{-\kappa r_{ij}},
    \label{eq:yukawa}
\end{equation}
where $A_{ij}$ is a constant pre-factor and the Debye-length $\kappa$ is given by
\begin{equation*}
    \kappa = \sqrt{2 e^2 k_B T \epsilon_0 \epsilon_r(T) c_s},
\end{equation*}
where $c_s$ is the ionic strength, $e$ the electron charge, $k_B$ is Boltzmann's constant, $\epsilon_0$ is the vacuum dielectric constant and $\epsilon_r$ is the relative dielectric constant of the implicit water solution computed according to an empirical relation~\cite{akerlofDielectricConstantWater1950} (See Sec.~IVA in the SI for details).

Due to the appearance of the Debye-length $\kappa$, Eq.~\eqref{eq:yukawa} mirrors Debye-H\"uckel theory, which indeed can be recovered exactly if $A_{ij} \propto q_i q_j$, where $q_{i/j}$ are the charges of the interacting residues. Instead of deriving the pre-factor $A_{ij}$ strictly from the charges $q_i$ and $q_j$, the Mpipi-Recharged model utilizes a small pair-dependent deviation to account for asymmetries observed in atomistic potential of mean force calculations~\cite{feito_2026_determination}.

\subsubsection{Setup details}

Each evaluation of the total objective function Eq.~\eqref{eq:obj_tc_total} requires the evaluation of a density at four temperatures. These densities were obtained by running small NPT simulations of 27 proteins for 100 ns. After equilibration, the density was averaged over the last 10 ns of the trajectory, with a total of 1000 samples. All simulations use an ionic strength of 150 mM --- close to the salt concentration at physiological conditions. 

We made use of all three modes of parallelism discussed in Sec.~\ref{subsec:concurrency}:
\emph{Simulation engine parallelism} was utilized by running LAMMPS with 16 cores per temperature. \emph{Objective function parallelism} was used to parallelize the objective function over temperature points (see Sec.~IVC in the SI for details of the LAMMPS simulations), resulting in 64 used cores per objective function evaluation. Lastly, we made use of \emph{parameter trial parallelism} by evaluating two parameter sets concurrently on each cluster node over a total of 10 nodes, leading to 20 parameter sets being evaluated in parallel for each step.

We used Nevergrad's ``NgIohTuned" meta-optimizer~\cite{nevergrad} with 30 ``steps", leading to a total of 600 objective function evaluations. The wall time was 96 hours.

Details of the DC simulations run for validation are provided in Sec.~IVC of the SI. Error bars in the densities of the dense and dilute phases were estimated by bootstrapping, dividing the last $100$ samples of the profile into batches of $25$ each.

%%%%%%%%%%%%%%%%%%%%%%%%%%%%%%%%%%%%%%%%%%%%%%%%%%%%%%%%%%%%%%%%%%%%%
%% The "Acknowledgement" section can be given in all manuscript
%% classes.  This should be given within the "acknowledgement"
%% environment, which will make the correct section or running title.
%%%%%%%%%%%%%%%%%%%%%%%%%%%%%%%%%%%%%%%%%%%%%%%%%%%%%%%%%%%%%%%%%%%%%

\section*{Acknowledgements}
A.G. acknowledges a postdoctoral grant from the Icelandic Research Fund (grant no. 228615).
E.O.J acknowledges a project grant from the Icelandic Research Fund (grant no. 2410644).
J. R. E. acknowledges funding from the University of Cambridge, the Ramon y Cajal fellowship (RYC2021-030937-I), the Spanish scientific plan and committee for research reference PID2022-136919NA-C33, and the European Research Council (ERC) under the European Union’s Horizon Europe research and innovation program (grant agreement no. 101160499).
The calculations were performed using compute resources provided by Barcelona Supercomputing Center - Centro Nacional de Supercomputación (The Spanish National Supercomputing Center), the Icelandic Research Electronic Infrastructure (IREI) and the ARCHER2 UK National Supercomputing Service (\url{https://www.archer2.ac.uk})~\cite{beckettARCHER2ServiceDescription2024}.
We are grateful to Magnus A. H. Christiansen for providing the DFT calculations which were used as a reference for fitting the SCME/f parameters.
We thank Eduardo Pedraza Granado for sharing data and providing valuable advice on the direct coexistence simulations.
We thank Carlos Vega and Rohit Goswami for fruitful discussions.

%%%%%%%%%%%%%%%%%%%%%%%%%%%%%%%%%%%%%%%%%%%%%%%%%%%%%%%%%%%%%%%%%%%%%
%% The same is true for Supporting Information, which should use the
%% suppinfo environment.
%%%%%%%%%%%%%%%%%%%%%%%%%%%%%%%%%%%%%%%%%%%%%%%%%%%%%%%%%%%%%%%%%%%%%
\section*{Data Availability}

The authors confirm that the data supporting the findings of this study are available within the article and/or its supplementary materials.
The supporting information contains an overview of the backends used in the benchmark in Fig.~\ref{fig:chemfit_code}, a description of reduced units used and the simulation protocol for the density simulations in the LJ example, a plot comparing experimental densities of liquid argon with those obtained from simulations with the optimal LJ parameters, a figure depicting the LJ parameters found in this work compared to those from the literature, details about the electrostatic damping in the SCME and the optimization results for each of the eight parameters optimized, the definition of the temperature-dependent dielectric constant used in the Mpipi-Recharged, further details about the density-jump based objective function used in the CG model optimization, sequences of the WT and +7R variants of the LCD of hnRNPA1, details of the simulation protocol for the density simulations, details of the DC simulations used for validation and of the analysis of the density profiles obtained from DC, and Tab.~S4 contains all twenty-two optimized parameters along with the original parameters.
Snapshots in this work were created using \texttt{Solvis2}~\cite{solvis2_code, solvis_code}.

\section*{Code Availability}
\texttt{ChemFit} is a free and open-source code, available on GitHub~\cite{chemfit_code} with an online documentation~\cite{chemfit_docs}.
Code for running the examples in this article are also available on GitHub~\cite{chemfit_examples}.
Further, we provide a record in the Materials Cloud Archive with several input and output files~\cite{DataAvailabilityChemFit}, as well as snapshots of the source code of the SCME potential implementation and of ChemFit.

\putbib  % prints the main-text bibliography here

\end{bibunit}

%%%%%%%%%%%%%%%%%%%%%%%%%%%%%%%%%%%%%%%%%%%%%%%%%%%%%%%%%%%%%%%%%%%%%
%% SUPPORTING INFORMATION
%%%%%%%%%%%%%%%%%%%%%%%%%%%%%%%%%%%%%%%%%%%%%%%%%%%%%%%%%%%%%%%%%%%%%

\begin{bibunit}
\renewcommand{\bibnumfmt}[1]{S#1.}
\renewcommand{\citenumfont}[1]{S#1}
\setcounter{NAT@ctr}{0}

% Disable citation hyperlinks in the SI only.
% This avoids duplicate hyperref anchors from main/SI bibliographies.
\makeatletter
\renewcommand{\hyper@natlinkstart}[1]{}
\renewcommand{\hyper@natlinkend}{}
\makeatother

\clearpage
\onecolumngrid

\begin{center}
{\Large \textbf{Supporting Information}}\\[1em]
{\large ChemFit: A framework for automated high-dimensional model parameter optimization}\\[1em]
Moritz Sallermann,
Amrita Goswami,
Rosana Collepardo-Guevara,
Alberto Ocana,
Hannes J\'{o}nsson,
Elvar \"O. J\'{o}nsson,
Jorge R. Espinosa
\end{center}

\FloatBarrier

\vspace{1cm}

% S-numbering...
\setcounter{section}{0}
\setcounter{subsection}{0}
\setcounter{subsubsection}{0}
\setcounter{figure}{0}
\setcounter{table}{0}
\setcounter{equation}{0}

\renewcommand{\thesection}{\Roman{section}}
\renewcommand{\thefigure}{S\arabic{figure}}
\renewcommand{\thetable}{S\arabic{table}}
\renewcommand{\theequation}{S\arabic{equation}}

\renewcommand{\theHsection}{SI.\arabic{section}}
\renewcommand{\theHsubsection}{SI.\arabic{section}.\arabic{subsection}}
\renewcommand{\theHsubsubsection}{SI.\arabic{section}.\arabic{subsection}.\arabic{subsubsection}}
\renewcommand{\theHfigure}{SI.\arabic{figure}}
\renewcommand{\theHtable}{SI.\arabic{table}}
\renewcommand{\theHequation}{SI.\arabic{equation}}

%%%%%%%%%%%%%%%%%%%%%%%%%%%%%%%%%%%%%%%%%%%%%%%%%%%%%%%%%%%%%%%%%%%%%
%% Start the main part of the supplementary here.
%%%%%%%%%%%%%%%%%%%%%%%%%%%%%%%%%%%%%%%%%%%%%%%%%%%%%%%%%%%%%%%%%%%%%
\section{ChemFit code}

An overview of characteristics of the backends used in Fig.~1A of the main text is given in Tab.~\ref{tab:backends}.

\begin{table}
    \caption{Characteristics of backends used in Fig. 1A of the main text.}
    \label{tab:backends}
    \begin{tabular}{p{3cm}p{2cm}p{2cm}p{7cm}}
            \toprule
            Name &  & Multi node & Note\\ 
            \midrule
            Threadpool & Thread & No & {Low overhead, limited use for pure Python functions due to global interpreter lock (GIL)}\\
            Processpool & Process & No & {Medium overhead. Circumvents GIL.}\\
            Dask & Both & Yes & {High overhead. Extremely flexible.}\\
            MPI (ChemFit)& Process & Yes & {Low overhead, less flexibilty. Only available to parallelize \textit{CombinedObjectiveFunction}. Requires launching multiple instances of the script, which needs special considerations.}\\
    \bottomrule
    \end{tabular}
\end{table}

\section{Lennard-Jones example}
\label{section:lj}

In the Lennard-Jones (LJ) example, we use reduced units for the pressure and temperature, defined as: 

\begin{equation}
    \rho_\text{red} \equiv 
    \frac{\rho}{ \rho^*}
    =\frac{n_\text{atoms}}{V} \sigma^3
    ,
    \label{eq:reduced_density}
\end{equation}
where $\rho$ is the number density, $\rho^* = \sigma^{-3}$, $n_\text{atoms}$ is the number of atoms, and $V$ is the volume. The reduced temperature, $T_\text{red}$, is
\begin{equation}
    T_\text{red} \equiv \frac{T}{ T^*}
    = \frac{k_{\mathrm{B}} T}{\varepsilon},
    \label{eq:reduced_temperature}
\end{equation}
where $T$ is the temperature, $T^* = \frac{\varepsilon}{k_{\mathrm{B}}}$ and $k_{\mathrm{B}}$ is the Boltzmann constant.

\subsection{Simulation protocol}
\label{subsection:sim_method}

We used 1000 Argon atoms, with a  Nosé–Hoover thermostat~\cite{Evans1985} with a coupling constant of 100 fs, as well as the Nosé–Hoover barostat~\cite{Shinoda2004} with a coupling constant of 1000 fs.
For the initial box, we used a cube with a side length of
\begin{equation}
    L_\text{init} = 2 \sigma \left( n_\text{atoms} \right)^{1/3},
\end{equation}
in which we let LAMMPS randomly place atoms while maintaining a minimum distance of
\begin{equation}
    \Delta_\text{min} = 1.2 \sigma
\end{equation}
between atoms.
We use the "truncated and shifted" version of the LJ potential (\texttt{lj/cut} pairstyle in LAMMPS) with a cutoff radius of
\begin{equation}
    r_c = 5 \sigma.
\end{equation}

For each data point, the system was equilibrated for 2 ps in the NVT ensemble maintaining the initial box size $L_\text{init}$ and at the experimental temperature. Then, the system was equilibrated for 20 ps in the NPT ensemble at the experimental temperature and pressure. Lastly, the "production" run lasted another 20 ps and the density was computed by averaging the instantaneous densities every 2 ps. The timestep for all runs was 2 fs.

\subsection{LJ parameterization results}

\begin{figure}[h]
    \includegraphics[width=0.6\linewidth]{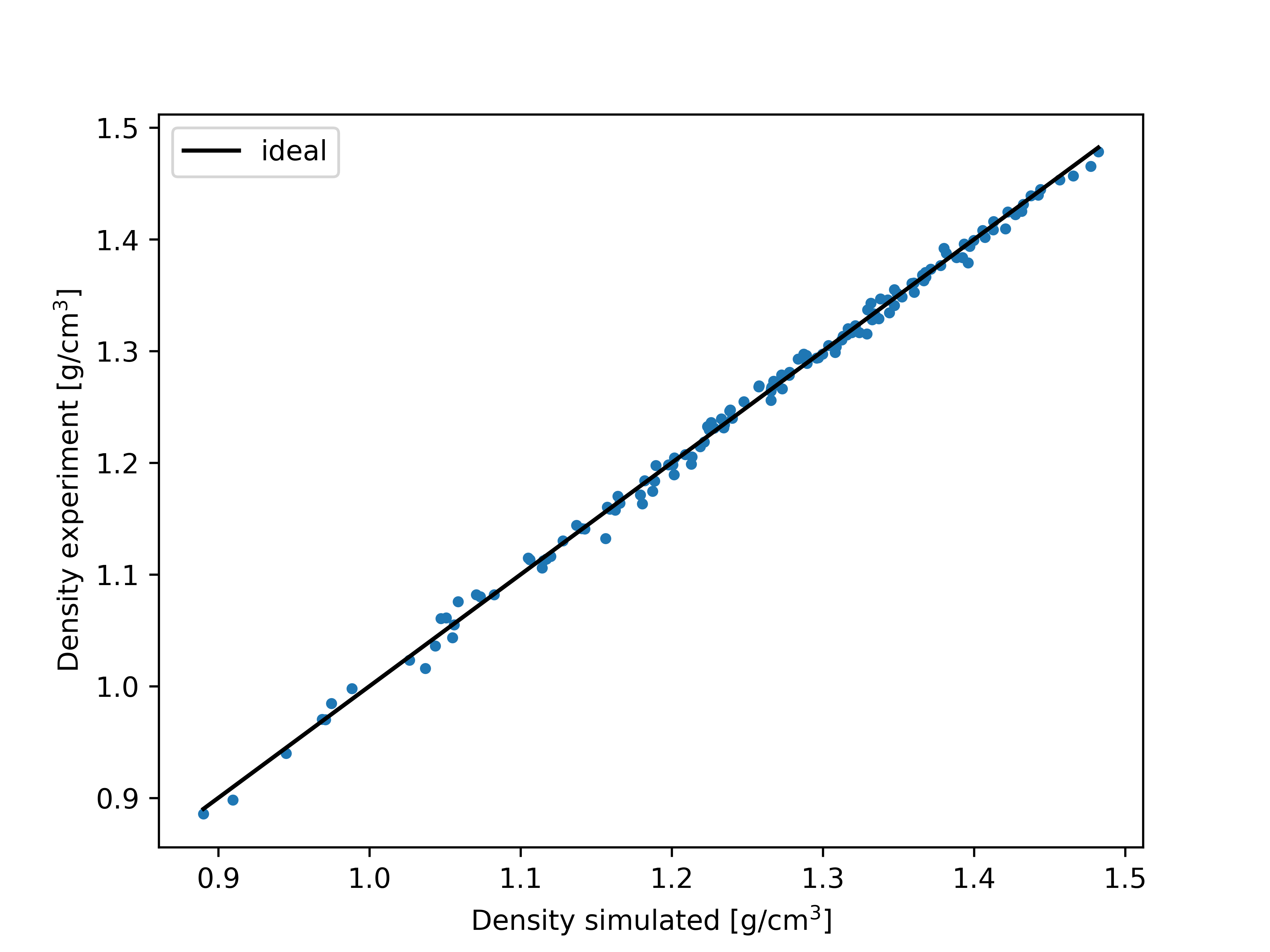}
    \caption{Density from the reference experiments~\cite{streett_1969_experimental} versus the density obtained from simulations using optimal parameters. The solid black line denotes the ideal scenario in which the simulated densities are identical to the reference densities. 
    }
    \label{fig:densities}
\end{figure}

Fig.~\ref{fig:densities} visually depicts the agreement of densities obtained from simulations of the LJ potential, using the optimal parameters, with the reference densities from experiments.~\cite{streett_1969_experimental}

\begin{figure}[h]
    \includegraphics[width=0.6\linewidth]{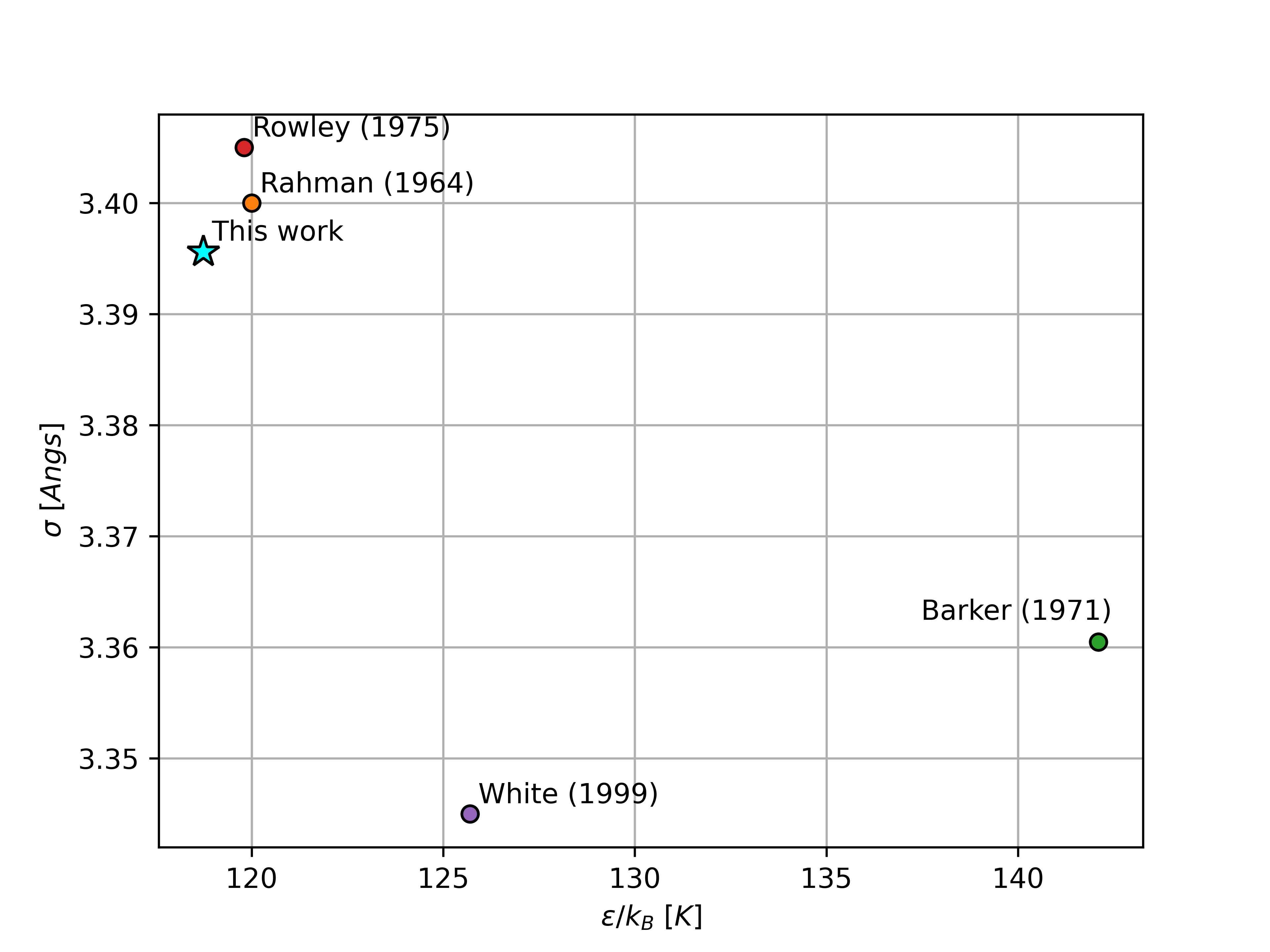}
    \caption{The size parameter, $\sigma$ (in \AA), versus the binding energy, $\varepsilon$ (in $k_\text{{B}}$), for the optimal parameters obtained in this work and those from the literature.~\cite{Rahman1964, Rowley1975, White1999, Barker1971}   
    }
    \label{fig:literature_values}
\end{figure}

Fig.~\ref{fig:literature_values} shows the optimal parameters (cyan star) obtained in this work, compared to those from the literature~\cite{Rahman1964, Rowley1975, White1999, Barker1971} in the $\sigma$-$\varepsilon$ space.

\section{SCME example}

\subsection{SCME electrostatic damping}

In general the electric field $V^{ij}$ at a position $r_i$ due to a moment $M$ at a position $r_j$ is given by an equation of the form
\begin{equation}
    V^{ij}_{\alpha} = T^{ij}_{\alpha \beta ... \gamma} M_{\beta...\gamma},
\end{equation}
where the notation $\beta ... \gamma$ means there are as many indices as the rank of the moment (i.e., two for a dipole moment, three for a quadrupole moment etc.) and 
$T^{ij}_{\alpha \beta ... \gamma}$ is a Coulomb tensor of appropriate rank.
The Coulomb tensors can be defined recursively as
\begin{align}
    T^{ij} &= \frac{1}{r} \lambda_0(r)\\
    T^{ij}_\alpha &= \nabla_{\alpha} T^{ij} = -\frac{r_\alpha}{r^3}\lambda_1(r)\\
    T^{ij}_{\alpha\beta} &= \nabla_{\beta} T^{ij}_{\alpha} = 3\frac{r_\alpha r_\beta}{r^5} \lambda_2(r) - \frac{\delta_{\alpha\beta}}{r^3}\lambda_1(r)
\end{align}
and so on.

The $\lambda_n(r)$ damping functions are due to \citet{stone_2011_electrostatica}:
\begin{equation}
    \lambda_n(r) = \lambda_n\Big(S(r)\Big) = \text{erf}(S) - \frac{2}{\sqrt{\pi}} e^{-S^2} \left( \sum_{i=1}^{n} \frac{2^{i-1} S^{2i - 1}}{(2i-1)!!} \right),
\end{equation}
where
\begin{equation*}
    S = \frac{r}{\tau_E},
\end{equation*}
and $x!! := 1 \cdot 3 \cdot ... \cdot (x-2) \cdot x$ denotes the double factorial.

\subsection{Optimization results}

Initial, optimized and literature parameters of the SCME potential are listed in Tab.~\ref{tab:scme_results}.

\begin{table*}[t!]
    \caption{Results of parameter optimization of the SCME potential in atomic units -- $E_\text{h}$ denotes one Hartree of energy and $a_0$ is the Bohr radius. The initial values, the values after optimization and the literature values according to \citet{jonsson_2022_transferable} are listed.}
    \label{tab:scme_results}
    \begin{tabular}{llrrr}
        \toprule
        \multicolumn{2}{l}{Parameter} & Initial value &  Optimal value & \citet{jonsson_2022_transferable} \\
        \midrule
        $\tau_E$ & [$a_0$]& $2.0$ & $2.012$ & $2.087$\\
        $\tau_D$ & [${a_0}^{-1}$] & $4.7$ & $4.21$ & $4.0$\\
        $C_6$ & [$E_\text{h}\, {a_0}^6$] & $46.44$ & $45.85$ &  $46.44$ \\
        $C_8$ & [$E_\text{h}\, {a_0}^8$] & $1141.7$ & $1141.75$ & $1141.7$ \\
        $C_{10}$ & [$E_\text{h}\, {a_0}^{10}$] & $33441$ & $33441$ & $33441$\\
        $A$ & [$E_\text{h}$] & $353.57$ & $1327.09$ & $299.493$\\
        $B$ & [] & $-0.146$ & $-1.14$ & $-0.5515$\\
        $C$ & [${a_0}^{-1}$] & $-2.002$ & $-1.986$ & $-1.836$\\
        \bottomrule
    \end{tabular}
\end{table*}

\section{Coarse-grained protein model example}

\subsection{Mpipi-Recharged model details}

The dielectric constant is determined according to the following empirical relation~\cite{akerlofDielectricConstantWater1950}
\begin{equation*}
    \begin{split}
        \epsilon_r(T) &= \frac{5321}{T} + 233.76 - 0.9297\ T\\
        &+ 1.417 \cdot 10^{-3}\ T^2 - 8.292 \cdot 10^{-7}\ T^3\\
        &\approx 78 \text{ at room temperature}.
    \end{split}
\end{equation*}

\subsection{Density jump objective function}

The objective function was defined by Eqs. (8) and (9), of the main text. For clarity, we repeat those definitions here:

For each sequence, we define an objective $O_\text{seq}$ as
\begin{equation}
    O_\text{seq} = -\frac{\rho_\text{seq}(T^\text{seq}_\text{lo}) - \rho_\text{seq}(T^\text{seq}_\text{hi})}{\rho_\text{seq}(T^\text{seq}_\text{lo}) + \rho_\text{seq}(T^\text{seq}_\text{hi}) },
    \label{eq:obj_tc_si}
\end{equation}
with $\text{seq} \in [\text{WT},\text{+7R}]$, resulting in an overall objective function of
\begin{equation}
    O = O_\text{WT} + O_\text{+7R}.
    \label{eq:obj_tc_total_si}
\end{equation}

Note that the maximum and minimum value of $O_\text{seq}$ are $0$ and $-1$, respectively, provided that $\rho_\text{seq}(T^\text{seq}_\text{hi})$ is greater than $\rho_\text{seq}(T^\text{seq}_\text{lo})$. The minimum value of the overall objective function, $O$, is then $-2$. 

Let us first define the critical temperature of a specific parametrization of the model as $\left[T^\text{seq}_C \right]_{\text{model}}$.
Now, we can assert the following:
    \begin{itemize}
    \item If $T^\text{seq}_\text{lo} \leq \left[T^\text{seq}_C \right]_{\text{model}} \leq T^\text{seq}_\text{hi}$ this will lead to $\rho(T^\text{seq}_\text{lo}) \gg \rho(T^\text{seq}_\text{hi})$ and Eq.~\eqref{eq:obj_tc} yields a value close to $-1$.
    
    \item If $T^\text{seq}_\text{hi} < \left[T^\text{seq}_C \right]_{\text{model}}$, both densities will approximately lie on the dense branch of the coexistence curve of the model giving an intermediate value between 0 and -1. As the $\left[T^\text{seq}_C \right]_{\text{model}}$ approaches $\left[ T^\text{seq}_C\right]_\text{exp}$ from above, Eq.~\eqref{eq:obj_tc} will approach $-1$.
    
    \item If $\left[T^\text{seq}_C \right]_{\text{model}} < T^\text{seq}_\text{hi}$, both simulated temperatures are in a supercritical phase and the objective function can quickly become unreliable. In our example, such an issue does not occur since the model $\left[T^\text{seq}_C \right]_{\text{model}}$ is well above the experimental value.
\end{itemize}
    
A primary reason for supercritical temperatures leading to unreliable results is easily understood by considering the regime of $T \gg \left[T^\text{seq}_C \right]_{\text{model}}$ and approximating the equation of state by the ideal gas law -- in the sense of neglecting inter-protein interactions. We obtain 
\begin{equation*}
    \rho \propto \frac{P}{T}.
\end{equation*}

But since we compute the densities at $P=0$ bar, both numerator and denominator of Eq.~\eqref{eq:obj_tc} become zero, making it undefined. Numerically, this leads to wildly fluctuating and unreliable results. One could potentially alleviate this issue by considering a finite but small pressure instead.
On the other hand, an implicit solvent model like the Mpipi-Recharged is not expected to perform well above its critical temperature regardless.

\subsection{Simulation protocol}
\label{subsection:mpipi_sim_method}

In the following, we provide information about the sequences used, details of the simulations performed during the opimization with \texttt{ChemFit}, as well as those of the direct coexistence (DC) simulations performed for validation.

\subsubsection{Protein sequences}
\label{subsection:mpipi_seq}

We have simulated the low-complexity domain (LCD) of the wild-type HNRNPA1, along with its Arginine-rich +7R variant.\\[0.35cm]

\textbf{+7R variant:}
\begin{center}
\begin{minipage}{0.85\linewidth}
\seqsplit{GSMASASSSQRGRSGRGNFGGGRGGGFGGNDNFGRGGNFSGRGGFGGSRGGGRYGGSGDRYNGFGNDGRNFGGGGSYNFGNYNNQSSNFGPMKGGNFRGRSSGPYGRGGQYFAKPRNQGGYGGSSSSRSYGSGRRF}
\end{minipage}
\end{center}

\textbf{Wild type:}
\begin{center}
\begin{minipage}{0.85\linewidth}
\seqsplit{GSMASASSSQRGRSGSGNFGGGRGGGFGGNDNFGRGGNFSGRGGFGGSRGGGGYGGSGDGYNGFGNDGSNFGGGGSYNDFGNYNNQSSNFGPMKGGNFGGRSSGPYGGGGQYFAKPRNQGGYGGSSSSSSYGSGRRF}
\end{minipage}
\end{center}

\subsubsection{NPT simulations}
\label{subsection:mpipi_npt}

For each sequence, $27$ replicas were placed in a periodic simulation box. Following energy minimization, equilibration simulations were run in the $NVT$ ensemble, ramping up the timestep gradually from $0.1$ fs to $10$ fs.

Production runs were performed in the $NPT$ ensemble. The Langevin thermostat~\cite{brunger1984stochastic} was used for all simulations, with a relaxation time of $10$ ps. The pressure was maintained at $0$ atm using the Berendsen barostat~\cite{berendsen1984molecular} with a relaxation time of $100$ ps. The simulations were run for a total of $100$ ns, using the last $10$ ns for averaging the density.

\subsubsection{Direct coexistence simulations}
\label{subsection:mpipi_dc}

An elongated prismatic box is used for direct coexistence (DC) simulations, such that the high-density and low-density phases are separated by an interface. In our setup, the high-density phase is sandwiched by the low density phase on either side. The longer side of the box ($z$ dimension) is perpendicular to the interfaces. Naturally, since the box is periodic in all dimensions, there is one high-density phase and one low-density phase, separated by two interfaces.

In order to create the high-density phase for each sequence, we compressed $128$ replicas of the protein in the $NPT$ ensemble at very high pressure (from $5$ atm to $20$ atm) for $2$ ns. Subsequently, the elongated box was created by adding vacuum to either side of the dense phase (in the $z$ dimension), such that the overall density in the box was $\approx 0.1$ g/cm$^3$. The overall density of the box is the same as that used by \citet{pedrazaPredictingSaturationConcentrations2025a}. Importantly, the volume of the entire box is a consideration when comparing dilute densities with a CG model with an implicit solvent. 

Additionally, the box dimensions were chosen such that the shorter sides of the box (in the $x$ and $y$ dimensions) were greater than at least twice the radius of gyration of the protein in the system, in order to avoid self-interactions.

Starting from such an initial configuration, each DC simulation was run in the $NVT$ ensemble for $1.0-1.5$ $\mu$s. 
Similar to Sec.~\ref{subsection:mpipi_npt}, the Langevin thermostat~\cite{brunger1984stochastic} was used with a relaxation time of $10$ ps. 

\subsubsection{Density profile analysis}
\label{subsection:density_profile}

The density profile is saved every $2$ ns. This profile is centered around $0$ at every frame prior to analysis. An time-averaged density profile is subsequently computed by averaging the last $100$ profiles, and then fitted to the following relation~\cite{tesei2023improved}:

\begin{equation}
    \rho(z) = \frac{\rho_{\text{dens}} + \rho_{\text{dil}} }{2} - \frac{\rho_{\text{dens}} - \rho_{\text{dil}}}{2} \times \tanh \left( \frac{|z| - z_{DS}}{t} \right),
\end{equation}

where $\rho_{\text{dens}}$ and $\rho_{\text{dil}}$ are the densities of the dense and dilute protein phases, respectively, $z_{DS}$ is the location of the dividing surface, and $t$ is the thickness of the interfacial region. 

\subsection{Critical temperatures of hnRNPA1 mutants}
Tab.~\ref{tab:tc_comparison} shows critical temperatures estimated by \citet{pedrazaPredictingSaturationConcentrations2025a} using the Mpipi-recharged model, compared with critical temperatures visually estimated from experimental phase diagrams by \citet{bremerDecipheringHowNaturally2022}. Only mutant variants for which the experimental data allowed for such a visual estimation are included --- for instance the $-$4D variant was ignored because there were only three experimental data points.

\begin{table}
    \centering
    \caption{Estimated critical temperatures from the simulations by \citet{pedrazaPredictingSaturationConcentrations2025a} and the experiments \citet{bremerDecipheringHowNaturally2022}. All temperatures are in Kelvin. The temperature difference is compute as $\Delta = [T_C]_\text{exp} - [T_C]_\text{mpipi-r}$.
    }
    \label{tab:tc_comparison}
    \begin{tabular}{lccr}
        \toprule
        Name & $[T_C]_\text{exp}$ & $\quad[T_C]_\text{mpipi-r}$ & $\Delta$ \\
        \midrule
        \textbf{WT} & 337 & 340     & $\quad\mathbf{-3}$ \\
        \textbf{+7R} & 285 & 341    & $\quad\mathbf{-56}$ \\
        +7R+12D & 343 & 399         & $\quad -56$ \\
        $-$12F+12Y & 330 & 340        & $\quad -10$ \\
        +7Y$-$7F & 322 & 317          & $\quad 5$ \\
        $-$3R+3K & 311 & 306          & $\quad 5$ \\
        $-$9F+6Y & 299 & 310          & $\quad -11$ \\
        +7K+12D & 320 & 348         & $\quad -28$ \\
        \bottomrule
    \end{tabular}
\end{table}

\subsection{Optimization results}
\label{subsection:mpipi_results}

Tab.~\ref{tab:mpipi_comparison} shows the canonical Mpipi-Recharged parameters~\cite{R.Tejedor2025} and the new optimized parameters.
Note that not all parameters were optimized, since twenty-two parameters were chosen such that the relative frequency of residues in each sequence was considered.

\begin{table}[ht]
    \centering
    \caption{Comparison of initial parameters of the base Mpipi-Recharged model and the optmized paramters obtained with ChemFit.}
    \label{tab:mpipi_comparison}
    \begin{tabular}{lcc}
        \toprule
        Parameter & Mpipi base & Optimized \\
        \midrule
        $\varepsilon_\text{GG}$ & 0.17298 & 0.17846 \\
        $\varepsilon_\text{GS}$ & 0.14262 & 0.14381 \\
        $\varepsilon_\text{GR}$ & 0.33532 & 0.34192 \\
        $\varepsilon_\text{GN}$ & 0.25776 & 0.26054 \\
        $\varepsilon_\text{GF}$ & 0.39682 & 0.39207 \\
        $\varepsilon_\text{GY}$ & 0.41949 & 0.42292 \\
        $\varepsilon_\text{SS}$ & 0.11226 & 0.11228 \\
        $\varepsilon_\text{SR}$ & 0.30607 & 0.30343 \\
        $\varepsilon_\text{SN}$ & 0.22740 & 0.23063 \\
        $\varepsilon_\text{SF}$ & 0.36813 & 0.35798 \\
        $\varepsilon_\text{SY}$ & 0.39079 & 0.39354 \\
        $\varepsilon_\text{RR}$ & 0.15073 & 0.15440 \\
        $\varepsilon_\text{RN}$ & 0.41698 & 0.42023 \\
        $\varepsilon_\text{RF}$ & 0.71700 & 0.72210 \\
        $\varepsilon_\text{RY}$ & 0.80421 & 0.79370 \\
        $\varepsilon_\text{NN}$ & 0.34254 & 0.34775 \\
        $\varepsilon_\text{NF}$ & 0.47695 & 0.47106 \\
        $\varepsilon_\text{NY}$ & 0.49961 & 0.48873 \\
        $\varepsilon_\text{FF}$ & 0.60043 & 0.60311 \\
        $\varepsilon_\text{FY}$ & 0.62310 & 0.64224 \\
        $\varepsilon_\text{YY}$ & 0.64576 & 0.65285 \\
        $A_\text{RR}$ & 4.0000 & 4.0715 \\
        \bottomrule
    \end{tabular}
\end{table}

\FloatBarrier

%%%%%%%%%%%%%%%%%%%%%%%%%%%%%%%%%%%%%%%%%%%%%%%%%%%%%%%%%%%%%%%%%%%%%
%% The appropriate \bibliography command should be placed here.
%% Notice that the class file automatically sets \bibliographystyle
%% and also names the section correctly.
%%%%%%%%%%%%%%%%%%%%%%%%%%%%%%%%%%%%%%%%%%%%%%%%%%%%%%%%%%%%%%%%%%%%%
\putbib

\end{bibunit}

\end{document}